\documentstyle[12pt]{article}
\newcommand{\gsim}{\mbox{\raisebox{-1.0ex}
        {$\stackrel{\textstyle ~>~}{\textstyle \sim}$}}}
\newcommand{\lsim}{\mbox{\raisebox{-1.0ex}
        {$\stackrel{\textstyle ~<~}{\textstyle \sim}$}}}

\newcommand{\keV}{\mbox{keV}}
\newcommand{\eV}{\mbox{eV}}
\newcommand{\cm}{\mbox{cm}}

\newcommand{\erg}{\mbox{erg}}
\newcommand{\K}{\mbox{K}}
\newcommand{\km}{\mbox{km}}
\newcommand{\Mpc}{\mbox{Mpc}}
\newcommand{\eph}{\epsilon_{\gamma}}
\newcommand{\HI}{{\rm HI}}
\newcommand{\HII}{{\rm HII}}
\newcommand{\HeI}{{\rm HeI}}
\newcommand{\HeII}{{\rm HeII}}
\newcommand{\HeIII}{{\rm HeIII}}

\begin{document}

\baselineskip 7.5mm

\begin{flushright}
    ICRR-Report-301-93-13
    August, 1993
\end{flushright}

\vskip 1.5cm

\begin{center}
{\Large\bf Reheating  during Hierarchical Clustering
in the Universe Dominated by the Cold Dark Matter}

\vskip 0.8cm
{\large M. Fukugita}\\
{\it Yukawa Institute for Theoretical Physics, Kyoto University,
Kyoto 606}\\
{\large and}\\
{\large  M. Kawasaki} \\
{\it Institute for Cosmic Ray Research,  University of Tokyo,
Tanashi, Tokyo 188}\\
\end{center}

\vskip 1.5cm

\begin{center}
{\bf Abstract}
\end{center}

We investigate reheating of the universe by early formation of stars
and quasars in
the hierarchical clustering scheme of cold dark matter scenario,
with perturbation fluctuations normalized by the COBE data.
It is found that ionizing uv flux from OB stars with the abundance
given by the standard initial mass function is strong enough to
ionize the universe from $z\approx 30$ to the present epoch, if
1--2\% of the collapsed baryons go into stars.  This lessens
significantly the CMB anisotropies at a small angular scale.
Reionization also increases the Jeans mass to $M_{\rm luminous}\approx
10^9M_\odot$ for $z\lsim 10$, which leads to a cut-off of the
luminosity function of normal galaxies on a faint side.
A strong uv flux is expected at $z\approx 2-5$, and the null result
of the Gunn-Peterson test is naturally explained.
Early star formation also results in production of heavy elements,
and the observational metal abundance sets
a strong constraint on the photon energy injection into
the intergalactic space.


\vskip 0.5cm

\noindent
{\bf Key words}: cosmology:theory -- dark matter -- diffuse
radiation -- galaxies:formation

\newpage

\section{Introduction}

The currently most attractive scenario for cosmic structure formation
assumes the cold dark matter dominated universe and the initial
fluctuations originating from adiabatic Gaussian perturbations with
scale-invariant spectrum (Peebles 1982, Blumenthal et al.1984).
The fluctuations enter non-linear regime
at redshift $z\lsim 50$, depending upon the mass of the fluctuations
and the magnitude of each peak of the
Gaussian noise, and make gravitationally bound objects. A number of
N-body simulations have shown that this is indeed a successful
scenario for formation of large scale structure (e.g., Frenk { et
al.} 1988) and also of
galaxies (White \& Frenk 1992).
The evidence in support for this scenario
has been  given by the COBE-DMR experiment for anisotropies of cosmic
microwave background radiation (CMB); the data show that the
fluctuations just after the epoch of recombination for angular scales
larger than a few degree are close
to scale-invariant spectrum with the magnitude almost correctly
given by the CDM model of structure formation (Smoot et al. 1992).
It is also pointed out that the CDM dominated universe predicts
correctly the epoch and the abundance of bright quasars
with the assumption that
they were formed from high standard deviation peaks of the fluctuations
(Efstathiou \& Rees 1988).

At a more quantitative level the standard CDM model does not quite fit the
precise perturbation spectrum given by COBE-DMR and that from
correlation functions of galaxies (Efstathiou, Bond \& White 1992;
Davis, Summers \& Schlegel 1992), which motivates some authors to
consider modifications of the model (Adams et al. 1992; Davis,
Summers \& Schlegel 1992; Klypin et al. 1992; Efstathiou, Bond \& White 1992).
The original model, however,
seems to be at least a
good enough starting ground for a consideration of the history of
the universe.

In the CDM scenario structure formation is hierarchical. A small object
becomes a  gravitationally bound system first, and the mass of the bound
systems gradually increases by merging. In this scenario formation of
objects with a specific mass is not instantaneous, but statistical, and
the abundance of bound objects with a specific mass
varies as a function of redshift.

When bound
systems are formed, it is likely that a substantial fraction of baryons
form stars, if the cooling time
is shorter than the dynamical time.
The photons emitted from OB stars would then ionize the universe,
which
would affect formation and evolution of cosmic structure, as first
pointed out by Doroshkevich, Zeldovich and Novikov (1967)  and then discussed
by Couchman and Rees (1986).  In this paper we examine
detailed thermal history of the universe in the epoch at $z<100$
taking account of the effect of star formation
by explicitly solving the evolution equation for thermal history,
and discuss the consequences on formation of galaxies and background
photons in the universe.  We also take into account the uv flux from early
quasars expected in the CDM model.

Another interesting consequence of  early star formation is
that it results in production of heavy elements,
as needed from the metal
observation of globular clusters.  Actually the observed metal
abundance in population II objects sets a strong
constraint on the fraction of baryons which go into to stars, and
hence on the energy injected into the intergalactic space.

In our work we take the Press-Schechter theory (Press \&
Schechter 1974) for hierarchical
clustering, since it is known to describe well the clustering
observed in N-body simulations  (e.g., Efstathiou \& Rees 1988). We fix
the normalization of the Gaussian field by the data given from the
COBE-DMR experiment.  We confine ourselves to  a flat universe
with $\Omega=1$ for simplicity.  We assume that
a certain fraction of collapsed baryons become stars, in so far as
the cooling condition is satisfied.  We also
assume the initial mass function as observed in the Milky Way today
(Shapiro \& Teukolsky 1983), and adopt blackbody spectrum
with temperature specified by the temperature-mass relation for
population II stars.  Quasar formation is normalized to the
observed abundance at $z=2$.
The only free parameters of our calculation is
the fraction of baryons which form stars apart from the Hubble
constant.

The aspects which we pay a particular attention in the present
work are as follows:  Whether
the universe is reionized and whether it recombines again at low redshift
is of our
prime interest. If the universe would be reionized early enough
comparable to a unit Thomson optical depth, this
lessens anisotropies of CMB at an angular scale smaller than a
degree, which would make them consistent with the limit from the South
Pole II experiment (Gaier et al. 1992).  Ionization of the universe would
increase the Jeans mass, which otherwise is very small compared with
the galaxy mass today, to the order of irregular galaxies. The
exact Jeans mass limit depends on how efficient is ionization.   The evolution
of residual photons is also of significant interest.  For $z\lsim 5$
the Gunn-Peterson test (Gunn \& Peterson 1965; Steidel \& Sargent 1986;
Jenkins \& Ostriker 1991; Webb et al. 1992)
requires that the universe be ionized to a high
degree.  This means that a reasonable amount of
the uv flux must exist all the time for the relevant redshift range
to keep the universe ionized.  The necessity of uv flux
at around $z\approx 4-2$ is also motivated by the presence of
Lyman $\alpha$ clouds; the uv flux heats up the clouds so that they
do not collapse (Sargent
et al. 1980; Ostriker \& Ikeuchi 1983; Rees 1986).

The effect of reheating has been studied by a number of authors
(Couchman 1985,
Stebbins \& Silk 1986;  Couchman \& Rees 1986,
Bartlett \& Stebbins 1991; Barcons et al. 1991;
Fukugita \& Kawasaki 1990; 1993; Gnedin \& Ostriker 1992; see Cen et
al 1990; Cen \& Ostriker 1992 for a different class of the reheating
model) in
different contexts and models. In particular
Couchman \& Rees (1986) studied
qualitatively the possibility of reheating in the hierarchical
clustering scenario. We shall study more quantitatively
the thermal history of the
universe, with the strength of the fluctuations determined from the
CMB anisotropy observations.  Reheating in the hierarchical clustering
was also considered by  Sasaki et al. (1993)
However, these authors take account
only of the energy liberated from the gravitationally binding
objects. Such
scenario basically does not ionize the CDM universe.
The energy injection
from stars is crucial for reionization.

In sect.2  the basic model of the present calculation is described.
The method of the calculation for the thermal history is given in sect.3;
the thermal processes that we considered are described in detail. In
sect. 4 we present the results of our calculation and discuss the
meaning and implications of the results.
Sect.5 is given to a conclusion.

\section{The model}
\subsection{Press-Schechter theory}

In the Press-Schechter theory the comoving number density of non-linear
objects within the mass range between $M$ and $M+dM$ is given by
\begin{equation}
     N(z,M)dM = \sqrt{\frac{2}{\pi}}\frac{\rho_0}{M}
                \frac{\delta_c}{D_1(z)}
                \left(- \frac{1}{\sigma^2(M)}
                \frac{\partial \sigma}{\partial M}\right)
                \exp\left(
                 -\frac{\delta_c^2}{2\sigma^2(M)D_1^2(z)}\right),
\end{equation}
where $\rho_0$ is the mean comoving mass density, $\delta_c$ is the
overdensity threshold for the collapse ($=1.68$ corresponding to the
prediction of
spherical collapse model), $D_1(Z)$ is the Peebles function for the growth
of the perturbations ($ D_1(z) = (1+z)^{-1}$ for $\Omega= 1$), and
$\sigma^2(M)$ is the Gaussian variance given by
\begin{equation}
      \sigma^2(M)  =  \frac{1}{2\pi^2}\int_0^{\infty}
                  P(k)\exp(-r^2_M k^2)k^2 dk,
\end{equation}
with $r_M = (M/\rho_0)^{1/3}/\sqrt{2\pi}$  the width of the Gaussian
filter and $P(k)$  the power spectrum of density fluctuations. We
use the fitting formula for the CDM power-spectrum given by Bardeen,
Bond, Kaiser \& Szalay (1986):
\begin{eqnarray}
      P(k)   &= &  A |T(k)|^2 k,\label{spec}\\[1em]
         &  T(k) &  =  \frac{\ln (1+2.34q)}{2.34q}\nonumber\\[0.6em]
         & & \times
             [ 1 + 3.89q + (16.1q)^2 + (5.46q)^3 + (6.71q)^4]^{-1/4},\\[1em]
         & q  & \equiv  \frac{k}{\Omega h^2 {\rm Mpc}^{-1}},
\end{eqnarray}
where $h$ is the Hubble constant $ H_0$ in units of
$100\km\sec^{-1}\Mpc^{-1})$.  This gives a spectrum quite similar
to that given in Efstathiou, Bond \& White (1992), but with a slightly more
power for small scales.
The normalization $A$ of the spectrum is fixed by the COBE-DMR
data using
the relation (Efstathiou, Bond \& White 1992):
\begin{equation}
    \frac{Q_{\rm rms}}{T_0} = \left(\frac{5}{6\pi^2}\right)^{1/2}
                          \left(\frac{H_0}{2c}\right)^2
                          \Omega^{0.77} A^{1/2},
\end{equation}
where $T_0$ is the present temperature of CBR and $Q_{\rm rms}$ is the
fluctuation amplitude at $\sim 10^{\circ}$ (Smoot et al. 1992):
\begin{equation}
     Q_{\rm rms} = 16.7 \pm 4.6~ \mu \mbox{K}.
\end{equation}
This normalization fixes $N(M,z)$ now uniquely.

The fraction $\Omega_{\rm coll}$ of the
total mass of the universe which collapses into the bound object with
mass greater than $M$ is shown in Fig.1(a). The
effect of reionization is not taken into account here.
The Jeans mass before reionization
is given by (Bond \& Szalay 1983)
\begin{equation}
   M_{B,J} = 1.4\times 10^{5}M_{\odot}\Omega_B\Omega^{-3/2}h^{-1}
            (1+100/z)^{-3/2}
\end{equation}
which stays at a constant value $1.4\times 10^4\Omega^{-3/2}M_\odot$ for
$z\gsim 100$ due to residual ionization left over the recombination epoch.
The total mass corresponding to this Jeans mass is
$ 3\times 10^{5}\Omega^{-1/2}M_\odot$ (or less for a smaller $z$).

Another necessary condition for
continuing collapse is that the cooling time is shorter than the
dynamical time (Blumenthal et al. 1984),
\begin{equation}
    M_{B,c}> 3.7\times 10^5 (\Omega h^2)^{-0.917}
        \left(\frac{Y_e}{10^{-4}}\right)^{-0.625}
        \left(\frac{\Omega_B}{\Omega}\right)^{-1.04}
        \left(\frac{1 +z}{10}\right)^{-2.75}M_{\odot}
\end{equation}
where $Y_e\simeq 10^{-5}\Omega_B^{-1}\Omega^{1/2}h^{-1}$.
At $z=50$ the limiting mass is $M\sim 10^6M_\odot$ and decreases
as a function of redshift.  Therefore we conclude that the first objects
which collapse around $z\approx 50$ should have a mass of $10^6M_\odot$.

In the following calculation we take account of the effect of reionization
on the Jeans mass, which is written
\begin{equation}
   M_{B,J} = 1.4\times 10^{5}M_{\odot}\Omega_B\Omega^{-3/2}h^{-1}
            \left(\frac{T_e}{\mu T_{\gamma}}\right)^{3/2},
\end{equation}
where $T_e$ is the electron temperature, $T_\gamma$ is the temperature
of the cosmic background radiation and $\mu$ is the mean molecular
weight.
When the universe is reheated, $T_e$ becomes much larger than
$T_{\gamma}$ and $Y_e$ increases. Therefore, Jeans mass $M_{B,J}$
increases whereas $M_{B,c}$ decreases.
We take the object with max$[(\Omega/\Omega_B)M_{B,c},
(\Omega/\Omega_B)M_{B,J}] <M<10^{13}M_\odot$
as being collapsed. For $M>10^{13}M_\odot$ the cooling time becomes
longer than the Hubble time again, and galaxies are not formed
(Blumenthal et al. 1984).

\subsection{Early formation of stars}

When bound objects  are formed, we assume that a constant
fraction $f$ of the baryonic gas
goes into stars in so far as mass of the objects is greater than
the Jeans mass and smaller than the maximal mass of galaxies
($=10^{13}M_{\odot}$),
and the cooling time is shorter than the dynamical time.
We set $f=0$ otherwise.
We take the baryon density to be $\Omega_B=0.05 h^{-2}$ and
$\Omega_B/\Omega$ to be  universal.
We also assume
that the mass spectrum of stars is proportional to
the present-day initial
mass function (IMF), $\xi_{\rm imf}$, for which we adopt the
formula given by
Shapiro \& Teukolsky (1983). We take the
temperature-mass relation for population II stars
(e.g., Bond, Carr \& Hogan 1986):
\begin{equation}
    T_{\rm s}(M_{\rm s}) = 6\times 10^4~ \K~ \min\left[
                  \left(\frac{M_{\rm s}}{100M_{\odot}}\right)^{0.3}, 1
                  \right].
\end{equation}
Stars with mass smaller than $10M_{\odot}$ produce very few
ionizing photons.
Therefore it is sufficient if we only consider the uv flux
from stars whose mass is
greater than $10M_{\odot}$. A  star with
mass $M_{\rm s}$ produces the radiation energy $\epsilon_{\rm s} M_s$ in the
main-sequence time $t_{\rm MS}$, where $\epsilon_{\rm s}$ and $t_{\rm MS}$
are given by
\begin{equation}
    \epsilon_{\rm s}  =  0.0046~\left(\frac{X}{0.76}\right) \min \left[
                    \left(\frac{M_{\rm s}}{100M_{\odot}}\right)^{1/2},1
                    \right],
\end{equation}
and
\begin{equation}
    t_{\rm MS}  =  2.3\times 10^6 ~\mbox{year}
                \left(\frac{\epsilon_{\rm s}}{0.0046}\right)
                \max\left[ 1, \left(\frac{M_{\rm s}}{100M_{\odot}}\right)^{-2}
                \right],
\end{equation}
where $X=0.76$ is the hydrogen mass fraction.
Since the main-sequence time of massive stars with $M_s > 10M_{\odot}$
   is  shorter than the cosmic time for
 $z < 50$, we may assume that the uv photons
are produced instantaneously after the formation of the stars.
The production rate of uv photons is then
\begin{equation}
    \left(\frac{d n_\gamma(\epsilon_{\gamma})}{dz}\right)_{\rm star} =
          \int dM_{\rm s}  \frac{B(\eph,T_{\rm s})}{\eph}
          \epsilon_{\rm s} \xi_{\rm imf}(M_{\rm s})
          f\frac{\Omega_B}{\Omega}
          \int dM M \left(\frac{\partial N(M,z)}{\partial z}\right),
\end{equation}
where $B(\eph,T_{\rm s})$ is the blackbody spectrum
normalized to  $\int d\epsilon_{\gamma}B(\eph) = 1$.

Stars with mass greater than $4M_{\odot}$ eject heavy elements.
Following Carr, Bond \& Arnett (1984) we assume that
the fraction of mass ejected
as heavy elements is
\begin{eqnarray}
     Z_{\rm ej} & = & 0.5 -
                   \left(\frac{M_{\rm s}}{6.3M_{\odot}}\right)^{-1}
                 ~~~~~~~~ \mbox{for}~~ 15 < M/M_{\odot} < 100\\
     Z_{\rm ej} & = & 0.1
                 ~~~~~~~~~~~~~~~~~~~~~~~~~~~~
                          \mbox{for}~~~ 8 <M/M_{\odot}< 15 \\
     Z_{\rm ej} & = & 0.2
                 ~~~~~~~~~~~~~~~~~~~~~~~~~~~~
                          \mbox{for}~~~ 4 <M/M_{\odot}< 8.
\end{eqnarray}
Then evolution of the metallicity $Z$ is given by
\begin{equation}
    \frac{d Z}{dz} =
          \int dM_{\rm s}  Z_{\rm ej}\epsilon_{\rm s}
          \xi_{\rm imf}(M_{\rm s}) f\frac{\Omega_B}{\Omega}
          \int dM M \left(\frac{\partial N(M,z)}{\partial z}\right).
\end{equation}
In our work we take $f=0.02$ which, as we find, is about the upper
limit from the condition that metallicity of population II stars
does not exceed 10$^{-3}$.

We show in Fig.1 (b) the mass fraction of stars both with and without
taking the effect of reionization into account. The figure indicates
that a half of the stars are formed before
$z=10$. This is because we assumed that
stars are made always at a constant
fraction of collapsed baryons with the cut-off at
$M\leq 10^{13}M_\odot$.  In reality we should have  more stars
formed stationally from the collapsed gas, and
expect that our approximation would not
be good at low redshift, after galaxies are formed to their present
form.
Fig.1(c) shows the evolution of the emitted radiation energy from stars.
In principle, our scheme has a double counting problem in that the
baryons which have gone
into stars when a small object collapses might be used again
to make another stars when a larger object is formed.
Since the
fraction which goes into stars are very small ($\sim 10^{-2}$), however,
the double counting effect is actually negligible.

\subsection{Formation of quasars}

We take account of the effect of quasars following Efstathiou \&
Rees (1988).  Using the Press-Schechter theory, we estimate the  comoving
number
density of luminous quasars as
\begin{eqnarray}
    N_{\rm quasar}(z) &  = & \int_{\max [t-t_{\rm Q},0]}^t \int_{M_{\rm min}}
            \left(\frac{\partial N(M,z)}{\partial t}\right) dM dt\\[0.6em]
               &  \simeq & t_{\rm Q} \int_{M_{\rm min}}
            \left(\frac{\partial N(M,z)}{\partial t}\right) dM,
\end{eqnarray}
where $t_{\rm Q}$ is the lifetime of quasars which is assumed
to be much shorter
than cosmic time $t$, and $M_{\rm min}$ is the minimum mass of quasars. We take
$L_{\rm Q,min}=10^{47}$ erg sec$^{-1}$ as a cut-off luminosity of quasars
and estimate $M_{\rm min}$ from the mass-luminosity relation:
\begin{equation}
    M(L_{\rm Q}) = 2\times 10^{13}M_{\odot}
               \left(\frac{t_{\rm Q}}{10^8\mbox{year}}\right)
               \left(\frac{\epsilon_{\rm Q}}{0.1}\right)^{-1}
               \left(\frac{F_{\rm Q}}{10^{-4}}\right)^{-1}
               \left(\frac{L_{\rm Q}}{10^{47}\mbox{erg}}\right),
\end{equation}
where $F_{\rm Q}$ is the fraction of the collapsed matter
which becomes quasars,
and $\epsilon_{\rm Q}$ is the efficiency defined as the fraction of the
rest mass energy  that is converted  into radiation.
The quasar number density depends on $t_Q$ and $\epsilon_{\rm Q}
F_{\rm Q}$. We take
$t_{\rm Q} = 10^8$ year and fix $\epsilon_{\rm Q} F_{\rm Q}$ to be $10^{-5}$ to
give the observed comoving number density of quasars
brighter than $L_{\rm Q} = 2.5 \times 10^{46} h^{-2} \erg \sec^{-1}$ at $z=2$,
\begin{equation}
   N_{\rm quasar} \simeq 1.5 \times 10^{-8} h^{-3} \mbox{Mpc}^{-3}.
             \label{qso-dens}
\end{equation}

The uv photon production rate is given by
\begin{equation}
    \left(\frac{d n_\gamma (\epsilon_{\gamma},z)}{d t}\right)_{\rm quasar}
         = \frac{S_{\rm Q}(\epsilon_{\gamma})}{\epsilon_{\gamma}} \int
           t_{\rm Q} L_{\rm Q}
           \left(\frac{\partial N(M,z)}{\partial t}\right) dM,
\end{equation}
where $S_{\rm Q}(\epsilon_{\gamma})$ is the spectrum of the emitted photons
which we take to obey a power-law of the form (Miralda-Escud\'e \&
Ostriker 1990;
model QS2)
\begin{eqnarray}
     S_{\rm Q}  & \propto & \left\{
                    \begin{array}{ll}
                       \epsilon_{\gamma}^{-0.4} &
                           (\epsilon_{\gamma} < 10.2\eV) \\[1em]
                       \epsilon_{\gamma}^{-1.4} &
                           (\epsilon_{\gamma} > 10.2\eV )
                    \end{array}\right. ,
\end{eqnarray}
We normalize $S_{\rm Q}$ as $\int d\epsilon_{\gamma} S_{\rm Q} = 1$.

The abundance of quasars are given in Fig.1 (d).  The first quasars are
formed around $z\approx 7-8$. The comoving number density
reaches maximum at $z\approx 2.5$
and then it declines.
The radiation energy from quasars is shown in Fig.1(c).

\subsection{Heating by collapsing bound objects}

The possibility was discussed that liberation of kinetic energy
by radiation cooling during the collapse of galaxies
might ionize the universe (Hogan 1980).
If free-free cooling or recombination cooling is the dominant
cooling process, the produced photons might have energy high enough to
ionize the intergalactic medium. This possibility was reconsidered
within the context of hierarchical clustering theory recently by
Sasaki et al. (1993).
We can show, however, that this scenario does not work for the
following reasons:
for this scenario to work, free-free or
recombination cooling (we denote the cooling times as $\tau_{\rm ff}$
and $\tau_{\rm rec}$) must be faster than Compton cooling ($\tau_{\rm C}$),
expansion cooling ($\tau_H=1/H$) and line cooling ($\tau_{\rm line}$)
that follows collisional excitation.
The conditions $(\tau_{\rm ff}$ or $\tau_{\rm rec}) < (\tau_{\rm C}~
{\rm and}~ \tau_{\rm H})$
are satisfied only for low
temperature, $T < 2.5\times 10^5\K$, as seen in Fig.2
where the
critical temperature for $\tau_{\rm C}=\tau_{\rm ff}$ etc. are plotted as a
function  of redshift.
On the other hand, we can show
that, at such a low temperature, line cooling dominates over
recombination and free-free cooling (Fig.2).
This means that liberation of kinetic energy by radiation cooling
in binding objects is not efficient enough
and does not ionize the universe.  We give this argument
more quantitatively in Appendix A.

The situation gets worse if we include  line cooling due to helium.
On the other hand, the constraint
might be  weakened, if the inhomogeneity effect of baryon density
in collapsed objects that enhance recombination
(free-free) cooling is taken into account, as discussed by
Sasaki et al. (1993).
In any case, however, only a very narrow range of temperature may
be allowed to produce ionizing photons, even if this enhancement
is included.
The contributions from massive stars and quasars
are far more important in reionizing the universe.
We conclude that we can neglect the production of ionizing uv photons
from collapsing objects.

\section{Thermal history}

\subsection{Method of calculation}

The evolution equations are solved numerically
for the ionized fraction of hydrogen and helium
atoms $n({\rm HII})/n_{\rm H}$, $n({\rm HeII})/n_{\rm He}$
and $n({\rm HeIII})/n_{\rm He}$,
the photon spectrum $n_\gamma(\epsilon_\gamma,t)$, and the electron
temperature $T_e$ with all relevant thermal processes taken into
account.  The integration is made with the implicit Euler method.
The detailed expressions of thermal processes used
in our work are given in order in the following subsections.

We determine the spectral distortion of CMB by solving directly the
Kompaneets equation (e.g., Fukugita \& Kawasaki 1990).
Since reheating occurs at low $z$ and the temperature is not high
in the present model, the distortion is described by the
Zeldovich-Sunyaev spectrum and its magnitude is characterized well
(within the accuracy in solving numerically the Kompaneets equation)
in terms of the Compton $y_{\rm c}$-parameter:
\begin{equation}
    y_{\rm c} =\int dt \frac{k(T_e - T_{\gamma})}{m_e c^2}n_e\sigma_{\rm T} c.
\end{equation}

\subsection{Ionization of H and He}

The abundances of HI, HII, HeI, HeII, and HeIII are determined by the
balance between the ionizing processes and the recombination processes.
We treat a hydrogen atom as a two level system  $(1S,2S+2P)$ plus
continuum following Peebles(1968), Matsuda  et al. (1971)
and Jones \& Wyse (1985);
evolution of the HII fraction is determined by
\begin{eqnarray}
    \frac{d}{dt}\left(\frac{n(\HII)}{n_{\rm H}}\right) & = &
            \frac{R_{\rm 1c}n(\HI)}{n_{\rm H}}
             - \frac{\alpha_{2,\HII}n_e^2}{n_{\rm H}}\nonumber\\[0.6em]
    &+ &  \frac{R_{\rm 2c}n(\HI)}{n_H}~
          \frac{K( \alpha_{2,\HII} n_e^2
          +\Lambda n(\HI) e^{-h\nu_{\alpha}/kT_{\gamma}})
                +  e^{-h\nu_{\alpha}/kT_{\gamma}}}{1 + K (
                   \alpha_{2,\HII}n_e^2 + R_{\rm 2c}n(\HI) + \Lambda n(\HI))}
\end{eqnarray}
where $n_{\rm H} \equiv n(\HI)+n(\HII)$, $\Lambda (= 8.227\sec^{-1})$
is the two-photon decay rate from $2S$, $\nu_{\alpha}$ is the
Lyman-$\alpha$ frequency,  $K = c^3/(8\pi
\nu_{\alpha}^3)(a/\dot{a})$, $\alpha_{2,\HII}$ is the recombination
coefficient to $2S + 2P$ level and $R_{\rm 2c}$ ($R_{\rm 1c}$) is  the
ionization
coefficient from $2S + 2P$ ($1S$). $R_{\rm 2c}$ and $R_{\rm 1c}$ are given by
\begin{eqnarray}
    R_{\rm 2c} & = & \gamma_{\rm 2c} + \beta_{2,\HI} n_e
               + \int_{\epsilon_{2,\HI}}
                 c\sigma_{\rm 2f,\HI} n_{\gamma}(\eph)d\eph , \\[0.6em]
    R_{\rm 1c} & = & \beta_{1,\HI} n_e
               + \int_{\epsilon_{1,\HI}} c\sigma_{\rm 1f,\HI}
                  n_{\gamma}(\eph)d\eph ,
\end{eqnarray}
where
$\beta_{1,\HI}$ ($\beta_{2,\HI}$) is the coefficient for
collisional ionization from 1$S$ ($2S + 2P$) level,
$\gamma_{\rm 2f} = \alpha_{2,\HII}(T_{\gamma}) (2m_e kT_{\gamma})^{3/2}
e^{-3.4{\rm eV}/T_{\gamma}}$  is the photo-ionization coefficient
due to background
photons,  $\sigma_{\rm 1f,\HI}$ ($\sigma_{\rm 2f,\HI}$) is the
photo-ionization  cross
section from $1S$ ($2S+2P$)  and $n_{\gamma}(\eph)$ is the
spectrum of uv photons; we take $\epsilon_{1,\HI} =
13.6\eV$, $\epsilon_{2,\HI} = 10.2\eV$.
Explicit expressions for the coefficients used
in this work is summarized in Appendix B.

Since helium is a minor component, we may treat it  as a one-level system.
The
time evolution of HI, HII and HIII is then given by
\begin{eqnarray}
    \frac{d}{dt}\left(\frac{n(\HeII)}{n_{\rm He}}\right)
              & = & \frac{n(\HeI)}{n_{\rm He}}\int_{\epsilon_{\HeI}} d\eph
                      \sigma_{\rm bf,\HeI} c n_{\gamma}(\eph)
                      \nonumber\\[0.6em]
              & + &   \beta_{\HeI} n_e\frac{ n(\HeI)}{n_{\rm He}}
                      - \beta_{\HeII} n_e \frac{n(\HeII)}{n_{\rm He}}
                      \nonumber\\[0.6em]
              & - &   \alpha_{\HeII} n_e \frac{n(\HeII)}{n_{\rm He}}
                      + \alpha_{\HeIII} n_e \frac{n(\HeIII)}{n_{\rm He}}
                      \nonumber\\[0.6em]
              & - &   \xi_{\HeII} n_e \frac{n(\HeII)}{n_{\rm He}}, \\[1em]
    \frac{d}{dt}\left(\frac{ n(\HeIII)}{n_{\rm He}}\right)
               & = & \frac{n(\HeII)}{n_{\rm He}}\int_{\epsilon_{\HeII}} d\eph
                      \sigma_{\rm bf,\HeII} c n_{\gamma}(\eph)
                      \nonumber\\[0.6em]
              & + &   \beta_{\HeII} n_e \frac{n(\HeII)}{n_{\rm He}}
                      - \alpha_{\HeIII} n_e \frac{n(\HeIII)}{n_{\rm He}},
\end{eqnarray}
where $n_{\rm He}\equiv n(\HeI) + n(\HeII) + n(\HeIII)$, $\beta$, $\alpha$
and   $\xi$  are the coefficients for the collisional ionization,
recombination, and dielectronic recombination, respectively;
$\sigma_{\rm bf}$ is the
photoelectric ionization cross
section, $\epsilon_{\HeI}$ ($\epsilon_{\HeII}$) is
the ionization energy of HeI (HeII)
(see Appendix B).

\subsection{uv spectrum}

The time evolution of the photon spectrum is determined by solving
\begin{eqnarray}
   \frac{d n_{\gamma}(\eph)}{d t} &  = & \frac{\dot{a}}{a}
            \left(\eph\frac{\partial n_{\gamma}}{\partial \eph}
            - 2 n_{\gamma}\right) \nonumber\\[0.6em]
        &+ & \frac{1}{\eph}(j_{\rm ff} + j_{\rm fb,\HII} + j_{\rm fb,\HeII} +
             j_{\rm fb,\HeIII})\nonumber \\[0.6em]
        &- & (\sigma_{\rm 1f,\HI} + \sigma_{\rm 2f,\HI} + \sigma_{\rm bf,\HeI}
            + \sigma_{\rm bf,\HeII})c n_{\gamma} \nonumber \\[0.6em]
        &+ &  \left(\frac{dn_{\gamma}}{d t}\right)_{\rm star}
            + \left(\frac{dn_{\gamma}}{d t}\right)_{\rm quasar},
\end{eqnarray}
where $j_{\rm ff}$ is the emissivity of the free-free process,
$j_{\rm fb,\HII}$, $j_{\rm fb, \HeII}$ and $j_{\rm
fb,\HeIII}$ are that of the
free-bound process due to HII, HeII and HeIII, respectively.
In the present consideration the electron temperature is relatively low ($
\lsim 10^4\K$), and hence the uv photon production due to the free-free
process is not important.

The photons produced by recombination process have
a complicated spectrum. We treat HI and HeII as two
level systems $(1S, 2S+2P)$ and HeI as a three level system $(1^1S,
2^1S, 2^3P)$. When a free electron is captured by HII, HeII or HeIII,
the emmisivity is given by
\begin{equation}
  j_{{\rm fb},i,j} = \frac{\eph}{kT_e}\alpha_{i,j} n(i)n_e
               \exp(-(\eph - \epsilon_{i,j})/kT_e),
\end{equation}
where $i =$HII, HeII and HeIII, and $j$ represents the energy level;
$\alpha_{i,j}$ is the recombination coefficient and $\epsilon_{i,j}$ is
the binding energy of the $j$-level of an $i$-atom.
Unless an electron recombines directly into the 1$S$ level, a
monochromatic (line) photon or
two photons are subsequently emitted  from an excited atom at every
recombination process.
The photon emissivity is
\begin{equation}
  j_{{\rm fb},i,j,{\rm line}} =\eph \alpha_{i,j} n(i)n_e
                    \delta [\eph -(\epsilon_i -\epsilon_{i,j})],
\end{equation}
for a monochromatic photon emission, and
\begin{equation}
  j_{{\rm fb},i,j,{\rm two}} = 2 \eph \alpha_{i,j} n(i)n_e
                    \delta [\eph -(\epsilon_i -\epsilon_{i,j})/2],
\end{equation}
for the two-photon process.\footnote{Here we neglect the
effect of the energy distribution for the two-photon process.}

In case of dielectric recombination of HeII, two electrons are in
excited states after recombination, and they go down to the ground
state by emitting photons. We treat this process by assuming
that the
first electron goes to the ground state emitting a photon with energy
corresponding to $(2S+2P)$ level of HeIII and the second electron
emits a photon with energy corresponding to $2^1P$ level of HeII.
Therefore,
\begin{equation}
  j_{{\rm fb},\HeII,{\rm d}} = \eph  \xi_{\HeII} n(\HeII)n_e
               [\delta (\eph - 40.81\eV)+\delta (\eph - 21.2\eV)].
\end{equation}

%

\subsection{Electron temperature}

Evolution of the electron temperature is given by
\begin{eqnarray}
   \frac{3}{2}\frac{d}{dt} \left( \frac{kT_e n_B}{\mu}\right) & = &
                \sum_{i=\HI, \HeI, \HeII} n(i) c
                \int (\eph - \epsilon_i) n_{\gamma}
                \sigma_{{\rm bf},i} d\eph  \nonumber \\[0.6em]
         & - &  \sum_{i= \HI, \HeI, \HeII} \zeta_{i}n_e n(i)
                \nonumber \\[0.6em]
         & - &  \sum_{i= \HII, \HeII, \HeIII} \eta_{i}n_en(i)
                \nonumber\\[0.6em]
         & - &  \omega_{\HeII} n_e n(\HeIII)\nonumber\\[0.6em]
         & - &  \sum_{i= \HI, \HeI, \HeII} \psi_{i}n_e n(i)
                \nonumber \\[0.6em]
         & - &  \lambda_{\rm c} \nonumber \\[0.6em]
         & - &  \theta_{\rm ff} [n(\HII) + n(\HeII) + 4 n(\HeIII)]n_e
                \nonumber\\[0.6em]
         & - &  \frac{15}{2}\frac{\dot{a}}{a}
               \left(\frac{kT_e n_B}{\mu}\right),
               \label{Etemp}
\end{eqnarray}
where $\zeta_i$ is
the collisional-ionization cooling coefficient due to atomic
state $i$, $\eta_i$ is
the recombination cooling coefficient, $\omega_{\HeII}$ is the
dielectronic recombination cooling coefficient due to HeII, $\psi_i$ is the
collisional excitation cooling coefficient due to $i$, $\lambda_{\rm c}$ is
the Compton cooling rate and $\theta_{\rm ff}$ is the free-free cooling
coefficient (see Appendix).
The last term of (\ref{Etemp}) represents expansion
cooling. In the actual calculation we also include  Compton heating by
scattering of high energy photons off electrons, although it is not important
since the reheating occurs at low
redshifts ($z < 100$) and the energy of photons is small ($\lsim \keV$)
in our model.
Compton cooling, among other cooling processes,  is most important at $z
\gsim 5$; the electron temperature is basically determined
by the balance between Compton cooling and photoelectric ionization
heating.  At a lower redshift expansion cooling dominates.

\section{Results and implications}

The results of our calculation are presented in a few panels of Fig.3.
Fig. 3(a) shows a fraction of HI and HII;
the universe is reionized at $z\approx 30$, shortly
after the first objects with mass of $\sim 10^6
M_\odot$ collapse.  By this epoch the relative abundance of the
baryons collapsed into stars is $\Omega_{\rm star}/\Omega_B\simeq
10^{-5}$ and the photon energy density is
$\Omega_\gamma\simeq 10^{-10}$.  Since then ionization is kept at a
high degree; the HI fraction decreases below $10^{-6}$ by $z\simeq
20$ and stays at a value $<10^{-7}$  to
the present epoch, which explains the very strong limit on
neutral hydrogen from the Gunn-Peterson test (Gunn \& Peterson 1965;
Steidel and Sargent 1989; Jenkins \& Ostriker 1991; Webb et al 1992).
If we would switch off quasars, the hydrogen tends to recombine towards
$z\approx 1-2$  [$n(\HI)/n_{\rm H} \approx 10^{-4}$ at $z=0$].  However,
the uv flux
from quasars, which appears as a kink in the curve of HII around
$z\approx 6$ when they turn on, ensures continuous high
ionization to $z=0$.
We remark that the absorption due to possible Lyman $\alpha$ clouds
 at high $z$ might reduce the ionizing flux and lead to more HI
(Miralda-Escud\'e \& Ostriker 1992; Madau 1993),
which we ignore in the present work however.

Whether anisotropies with angular scale smaller than a few degrees
may be lessened in our reheating scenario is an interesting issue of
the model.
The Thomson optical depth calculated as
\begin{equation}
       \tau_{\rm T}=\int^\infty_0 dt c \sigma_{\rm T} n_e
\end{equation}
is $\tau_{\rm T} \simeq 0.22$ for the ionization history shown in
Fig.3(a) (the effect of helium is also taken into account).
While this optical depth is not so large, a  calculation made by
Sugiyama, Vittorio \& Silk (1993) indicates that the
Doppler peak of the Fourier
component of CMB correlation function $c_\ell$ at $\ell\simeq 200$
decreases by a factor 2 and also the power for $\sim 10^\prime$ by a
factor of 3 for reionization at $z\approx 30$.  Their
calculations also show that anisotropies at 1 degree scale
are appreciably smaller
than $1\times 10^{-5}$, and are perfectly allowed by the limit from
the South Pole II experiment (Gaier et al. 1992).

The Compton $y_{\rm c}$ parameter is also plotted in the same figure
(Fig.3(a)).  The
final value is $2\times 10^{-7}$.  This small $y_{\rm c}$ is understood by
the fact that ionization takes place at  low redshift
with low energy photons, which keep
electrons to stay at a low temperature; hence only a small
fraction of energy goes into
CMB by Compton cooling.  This contrasts the model of Gnedin \&
Ostriker (1991), where high energy photons are injected from accretion
activity of massive black holes which are formed at a very high
redshift; in their model $y_{\rm c}$ is as high as $10^{-4}$, though the
predicted value is perhaps flexible. We should also add a remark on
the difference of our model from the model with heating
by shock waves; the latter heats up electrons to 1-10keV and leads to
$y_{\rm c}$-parameter as large as $>
10^{-3}$ (Yoshioka \& Ikeuchi 1987); shock heating is allowed only at
small $z$ ( for this class of model, see Cen \& Ostriker 1992).

The metal abundance $Z$ (Fig.3(b)) closely follows the star abundance.  For
$z\lsim 10$, $Z$ stays at $(0.4-1)\times 10^{-3}$, which is a typical metal
abundance of population II stars.  This means that our fractional
star-formation parameter $f\simeq 2\times 10^{-2}$ is close to the limit;
namely, the observed metal abundance limits the amount of energy injection
from stars.

Ionization history of helium atoms is exhibited in Fig.3(c).  He I
is ionized to He II ($\Delta E= 24.6$eV) at the epoch of hydrogen
reionization, and $n({\rm HeII})/n_{\rm He}\simeq 10^{-5}$ after this
redshift.
He II is then fully ionized to He III ($\Delta E= 54.4$eV)
when quasars turn
on.
Since the starlight can not ionize He II, the
abundance of He III is taken as an indicator for the importance of
the quasar light (Mo, Miralda-Escud\'e \& Rees 1993).

In Fig.3(d) we plot the electron temperature, the baryon Jeans mass (10),
and the critical mass for the cooling condition (9).
We note that the electron temperature stays between $3\times
10^3 - 2\times 10^4$K for all period of $z<30$. The fact that
the universe is fully ionized despite this low temperature
can be understood by the slow recombination rate.

One of the most conspicuous effect in the reionized universe is a large
Jeans mass at low redshift (Couchman \& Rees 1986).  It
gradually increases from $10^4M_\odot$ for $z<100$ to
$1\times 10^9M_\odot$ for $z<6$. This means that normal galaxies
formed at $z<10$ should have
a mass function  with a peak
around $M_{\rm luminous}\approx 10^9-10^{10}$,
or a luminosity function with a peak at a $B$ magnitude
$M_B\approx -19$mag rather
than obey a Schechter function.
Observationally, little is known about the faint end of the
luminosity function of field galaxies. The only information available
on the faint end is from galaxies in the Virgo cluster; Binggeli,
Sandage \& Tammann (1985) obtained a complete galaxy sample down to
B=18mag, and
they have shown that the luminosity function of spiral
and elliptical galaxies are of the Gaussian type with peaks
at $M_B=-18.1$mag (dispersion $\sigma=1.5$mag) and $M_B=-18.2$mag
($\sigma=1.7$mag), respectively (Sandage, Binggeli \& Tammann 1985).
[Here we used $\langle (m-M)_0\rangle \simeq 31.4$ for spiral galaxies and
31.0 for elliptical galaxies; see the discussion in Fukugita, Okamura \&
Yasuda (1993).]
 The Jeans mass curve shown
in Fig.3(d) suggests that very small galaxies, such as dwarf
spheroidals, and globular clusters are very early objects which
collapsed at
$z>10$ and have survived merging to larger systems.

The critical cooling mass is smaller than the Jeans mass in the
reheated universe, in contrast to the case without reionization, for
which generally $M_{B,c} > M_{B,J}$.

The resulting photon spectrum is given in Fig.4 for $z=5, 3$ and 0.
The relative contributions from stars and quasars
are obvious in this figure. We show in Fig. 5 the strength of
ionizing flux  defined by
\begin{equation}
\bar J= \int_{\epsilon_{\HI}}^{\infty}
        \frac{d\epsilon_{\gamma}}{4\pi c}
        n_{\gamma}(\epsilon_{\gamma})
        \frac{\sigma_{\rm 1f,\HI}(\epsilon_{\gamma})}{
        \sigma_{\rm 1f,\HI}(\epsilon_{\HI})}
\end{equation}
as a function of redshift.  The ionizing flux rises very sharply
at $z\simeq 30$, the epoch of reionization.  As we noted earlier the
flux at low redshift may have substantial uncertainties, arising from
our underestimate of stationally star formation activity in normal
galaxies and from the neglect of the
absorption by Lyman $\alpha$ clouds.  Nevertheless, the value
of $\bar J$ around $z=5-3$ ($\bar J=(3-5)\times 10^{-21}$erg cm$^{-2}$
s$^{-1}$ sr$^{-1}$ Hz$^{-1}$) is interesting, since it is  about the
value required to keep Lyman $\alpha$ clouds from their collapse
(Sargent et al 1980; Ostriker \& Ikeuchi 1983; Rees 1986; Bajtlik,
Duncan \& Ostriker 1988).

\section{Conclusion and discussion}

We have shown that OB stars which would have been formed shortly
after the
collapse of very early objects, as expected in the standard CDM model,
ionize the universe completely at the epoch
as early as $z\approx 30$.  This
reionization significantly lessens the anisotropies of the CMB at an
angular scale smaller than 1$^\circ$ and removes the marginal
conflict between the COBE-DMR data and
the limit from the South Pole II experiment in the CDM model
(e.g., G\'orski, Stompor \& Juszkiewicz 1992), without invoking the aid
of the tensor perturbations (Crittenden, R. et al. 1993;
Davis, R.L. et al.  1992; Dolgov \& Silk 1992).
Reionization has also
an interesting implication, that luminosity function of normal galaxies
formed at $z<10$ should have a cutoff on the faint side,
which roughly corresponds to
the luminosity of dwarf galaxies. Smaller bound systems such as
globular clusters and dwarf spheroidals must be the
objects formed very early in the universe.

Another advantage of the present scenario is that the metal abundance
of globular clusters can be explained naturally.  The observed metal
abundance sets an upper limit on the photon energy that is emitted from stars
and injected into the universe.

We have shown that the null result of the Gunn-Peterson(GP) test is
satisfied by reheating in
contrast to the case of shock wave reheating as advocated by Cen
and Ostriker 1992; with shock heating alone ionization is not so
efficient and the
GP test is not satisfied by many orders
of magnitude.  The GP test is not satisfied in a heating model with
very early formation of massive black holes (Gnedin \& Ostriker 1992); the
light emitted from an early epoch is redshifted away and has no
ionizing power at low redshift. In order to keep hydrogen highly
ionized, the ionizing uv flux must exist during the relevant epochs.
This is satisfied in our model by successive formation of bound
objects.
We also
expect the ionizing flux  as strong as is required to
confine Lyman $\alpha$ clouds by continuous ionization.

In our reheating scenario the Compton $y_{\rm c}$
parameter takes a very
small value, much smaller than could be detected by the
COBE-FIRAS experiment (Mather et al. 1993).

The most interesting test for the present case is the CMB
anisotropies at a small angular scale.  If anisotropies
would be observed, for instance,  at a level of
$\Delta T/T\approx 1\times 10^{-5}$
reheating must not have happened before $z\approx 10-15$,
which means that
the baryon fraction that goes into stars should be smaller than
$f\simeq 10^{-3}$ in our terminology.

All of our calculation, except for the normalization of the fluctuations,
scales as $\Omega h^2$ (and $\Omega_B h^2$), and the normalization
of the  fluctuations depends very weakly on
$\Omega$ ($\Omega^{-0.2}$). Therefore, the reheating scenario changes
little if high value of the Hubble constant is taken in a low density
universe. If $\Omega h^2 \ll 1$, however, the epoch of reheating is
considerably delayed.

Our calculation presented here is certainly far from complete for
thermal history and evolution of galaxies after they are formed.
We have to carry out a calculation with a more astronomical treatment
to take proper account of the effects.

\vskip1cm
\noindent
{\bf Acknowledgement}

We would like to thank Martin Rees for his valuable comments.
\newpage

\appendix
\section{Quantitative argument for heating by
collapsing bound objects}

When the matter collapses, baryon gas and dark
matter particles are virialized and
distributed isothermally with radius $r_{\rm vir}$ and temperature
$T_{\rm vir}$, which are given by
\begin{eqnarray}
  r_{\rm vir} & = & \left[ \frac{M}{6\pi^3 \rho_{\rm vir}}\right]^{1/3}
                \nonumber\\[0.6em]
          & = & 1.7\times 10^{-5} \Mpc
                \left(\frac{M}{M_{\odot}}\right)^{1/3} (1+z)^{-1}
                (\Omega h^2)^{-1/3},
\end{eqnarray}
and
\begin{eqnarray}
   T_{\rm vir} & = & \frac{G m_p M \mu}{3 r_{\rm vir}}\nonumber\\[0.6em]
           & = &       1.04 \times 10^{-2} \K ~\mu
                \left(\frac{M}{M_{\odot}}\right)^{2/3} (1+z)
                (\Omega h^2)^{1/3},
\end{eqnarray}
where $\rho_{\rm vir}$ is the mean mass density of the virialized object and we
take $\rho_{\rm vir} = 180\rho$ from the spherical
collapse model.
The temperature of the baryon gas decreases by radiative cooling
processes, which convert the kinetic energy of baryon gas into
radiation.  If free-free cooling or recombination cooling dominates over
other cooling processes, the produced photons could have  energy high enough to
ionize the intergalactic medium.

This scenario requires that free-free or
recombination cooling  must be more efficient than Compton
cooling, expansion cooling and
line cooling.  For fully ionized plasma, the cooling times for the
free-free and recombination processes are
\begin{eqnarray}
     \tau_{\rm ff} & = & \frac{3T_e n_B}{2\mu \Lambda_{\rm ff}}
                     \nonumber\\[0.6em]
               & \simeq & 1.66\times 10^{14}\sec~ (T/\K)^{1/2}
                 (1+z)^{-3} (\Omega_B h^2)^{-1},
\end{eqnarray}
and
\begin{eqnarray}
     \tau_{\rm rec} & = & \frac{3T_e n_B}{2\mu \Lambda_{\rm rec}}
                     \nonumber\\[0.6em]
                & \simeq & 1.19 \times 10^{13}\sec~
                    \left(\frac{T}{\K}\right)^{0.7}
                    \left[1+\left(\frac{T}{10^6\K}\right)^{0.7}\right]
                    (1+z)^{-3} (\Omega_B h^2)^{-1},
\end{eqnarray}
where the cooling rates $\Lambda_{\rm ff}$ and $\Lambda_{\rm rec}$  are
given by
\begin{eqnarray}
    \Lambda_{\rm ff} & = & 1.42\times 10^{-27} g_{\rm ff} (T_e/\K)^{1/2}
                  (n(\HII)+4 n(\HeIII))n_e
                  ~\erg~\cm^{-3}~\sec^{-1}, \\[1em]
    \Lambda_{\rm rec}  & = & 6.5\times 10^{-27}
                  \left (\frac{T_e}{\K}\right)^{1/2}
                  \left(\frac{T_e}{10^3\K}\right)^{-0.2}
                  \left[1+\left(\frac{T_e}{10^6\K}\right)^{0.7}
                  \right]^{-1}\nonumber\\
                 &   & \times n(\HII) n_e~ \erg~\cm^{-3}~\sec^{-1},
\end{eqnarray}
with $n_e$, $n(\HII)$ and
$n(\HeIII)$ the number densities of electron,
HII and HeIII, and $g_{\rm ff}$ the gaunt factor of $\sim O(1)$.
That the cooling
time  $\tau_{\rm ff}$ or $\tau_{\rm rec}$ be shorter than the
Compton cooling time $\tau_{\rm C} = 3.9\times 10^{19}
(1+z)^{-4} \sec$ leads to the conditions:
\begin{eqnarray}
    \frac{\tau_{\rm C}}{\tau_{\rm ff}} & = & 29
                     \left(\frac{T_e}{10^4\K}\right)^{-1/2}
                     (1+z)^{-1}
                     \left(\frac{\Omega_B h^2}{0.0125}\right)
                     > 1, \label{tctff}\\[1em]
    \frac{\tau_{\rm C}}{\tau_{\rm rec}} & = & 65
                     \left(\frac{T_e}{10^4\K}\right)^{-0.7}
                     \left[1+\left(\frac{T_e}{10^6\K}\right)^{0.7}
                     \right]^{-1}
                     (1+z)^{-1}
                     \left(\frac{\Omega_B h^2}{0.0125}\right) > 1.
\end{eqnarray}
The cooling times $\tau_{\rm ff}$ or $\tau_{\rm rec}$ should also be
shorter than the dynamical time $\tau_H=1/H=(\dot a/a)^{-1}$
to make the cooling processes effective.
\begin{eqnarray}
     \frac{H^{-1}}{\tau_{\rm ff}} & = & 0.45
                  \left(\frac{T_e}{10^4\K}\right)^{-1/2}
                  (1+z)^{3/2}
                  \left(\frac{\Omega_B h^2}{0.0125}\right)
                  \left(\frac{\Omega h^2}{0.25}\right)^{-1/2} > 1,
                  \\[1em]
     \frac{H^{-1}}{\tau_{\rm rec}} & = & 0.99
                  \left(\frac{T_e}{10^4\K}\right)^{-0.7}
                  \left[1+\left(\frac{T_e}{10^6\K}\right)^{0.7}\right]^{-1}
                  (1+z)^{3/2}\nonumber\\[0.6em]
              & & \times \left(\frac{\Omega_B h^2}{0.0125}\right)
                  \left(\frac{\Omega h^2}{0.25}\right)^{-1/2} > 1.
                  \label{htrec}
\end{eqnarray}
These condition (\ref{tctff})-(\ref{htrec}) are satisfied simultaneously
for temperature $T < 3\times 10^5\K$, as seen in Fig.2, where the
critical temperature for $\tau_{\rm C}=\tau_{\rm ff}$ etc. are plotted as a
function  of redshift.

Another condition to produce efficiently ionizing uv photons
is that the recombination cooling
rate must dominate over the rate of line cooling of hydrogen atoms
that are excited collisionally.
The latter rate is given by
\begin{equation}
    \Lambda_{\rm line} = 7.5\times10^{-19}
                     \left[1+\left(\frac{T_e}{10^5\K}\right)^{1/2}
                     \right]^{-1}
                     \exp\left(-\frac{1.18\times 10^{5}}{T_e}\right)
                     n_e n(\HI)~ \erg~\cm^{-3}\sec^{-1},
\end{equation}
so that the ratio
$\Lambda_{\rm line}/\Lambda_{\rm rec}$ is
\begin{eqnarray}
    \frac{\Lambda_{\rm line}}{\Lambda_{\rm rec}} & = &
             2.9 \times 10^7
            \left(\frac{T_e}{\K}\right)^{-0.3}
            \left(1+\left(\frac{T_e}{10^5\K}\right)^{1/2}\right)^{-1}
            \nonumber\\[0.6em]
    & \times &
            \left(1+\left(\frac{T_e}{10^6\K}\right)^{0.7}\right)
            \frac{n(\HI)}{n(\HII)}
            \exp\left(\frac{-1.18\times 10^5\K}{T_e}\right).
\end{eqnarray}
The condition $\Lambda_{\rm line}/\Lambda_{\rm rec}<1$ means that
the fraction of HI be small enough.
Fig.6 shows the contour of $\Lambda_{\rm line}/\Lambda_{\rm rec}=1$
in the plane of $T$ versus
$[n(\HI)/n(\HII)]$; the production of photons requires that
the parameters should lie below this contour.
On the other hand, $n(\HI)/n(\HII)$ is determined by ionization
equilibrium, as
\begin{eqnarray}
   \frac{n(\HI)}{n(\HII)} & = & 4.95
            \left(\frac{T_e}{\K}\right)^{-1.2}
             \left[1+ \left(\frac{T_e}{10^5\K}\right)^{0.5}\right]
             \nonumber\\[0.6em]
    & \times &
             \left[1+ \left(\frac{T_e}{10^6\K}\right)^{0.7}\right]^{-1}
             \exp\left(\frac{1.58\times 10^5\K}{T_e}\right).
\end{eqnarray}
The curve showing the equilibrium condition is also plotted in Fig.6.
One can see that
the temperature should be greater than $3\times 10^5$K for
recombination cooling to dominate over line cooling.
However, this is just opposite to the condition for recombination
cooling being faster than Compton cooling.
This proves that liberation of energy in bound
objects does not efficiently produce ionizing photons.

\section{Ionization and recombination coefficients}

The ionization and recombination coefficients used
in our calculation is summarized in this Appendix.
The coefficients and cross sections used in our paper  are taken from
Black 1981;  Matsuda et al. 1971; Menzel \& Pekeris
1935; Spitzer 1978. We adopt the high temperature
correction given  by Cen 1992 for the coefficients of processes involving a
free electron and an orbital electron.

\subsection{Collisional ionization}

\noindent
(a) HI ($n=1 \rightarrow$ free):
\begin{equation}
   \beta_{1,\HI} = 5.85 \times 10^{-11} T^{1/2} (1 +
                   (T/10^5)^{0.5})^{-1}\exp(-1.578\times 10^5/T)
                   ~\cm^3 \sec^{-1},
\end{equation}
(b) HI ($n=2 \rightarrow$ free):
\begin{eqnarray}
   \beta_{2,\HI}& =& 5.46 \times 10^{-11} T^{1/2}
                     \exp(-3.882\times 10^4/T)\nonumber\\
                &\times & [19.98 - 5.89\times 10^{-5}T -2.81\times
                     10^4 T^{-1} + 5.44\times 10^7 T^{-2}]
                     ~\cm^3 \sec^{-1},
\end{eqnarray}
(c) HeI:
\begin{equation}
   \beta_{\HeI} = 2.38\times 10^{-11}T^{1/2}(1+(T/10^5)^{0.5})^{-1}
                  \exp(-2.853\times 10^5/T)
                   ~\cm^3 \sec^{-1},
\end{equation}
(d) HeII:
\begin{equation}
    \beta_{\HeII} = 5.68\times 10^{-12}T^{1/2} (1+(T/10^5)^{0.5})^{-1}
                  \exp(-6.315\times 10^5/T)
                   ~\cm^3 \sec^{-1}.
\end{equation}

\subsection{Recombination}

(a) HII (free $\rightarrow n\ge 1$):
\begin{equation}
    \alpha_{\HII} = 6.28\times 10^{-11}T^{-1/2}
                    \left(\frac{T}{10^3}\right)^{-0.2}
                   \left[ 1 +
                   \left(\frac{T}{10^5}\right)^{0.7}\right]^{-1}
                   ~\cm^3\sec^{-1},
\end{equation}
(b) HII (free $\rightarrow n\ge 2$) [Peebles 1993]:
\begin{equation}
    \alpha_{2,\HII} = 2.6\times 10^{-13}
                      \left(\frac{T}{10^4}\right)^{-0.8}
                      ~\cm^3 \sec^{-1},
\end{equation}
(c) HeII (free $\rightarrow n\ge 1$):
\begin{equation}
    \alpha_{\HeII} = 1.50\times 10^{-10}T^{-0.6353}~\cm^3 \sec^{-1},
\end{equation}
(d) HeII (free $\rightarrow n^3 [n\ge 2]$)
\begin{equation}
    \alpha_{\HeII,2^3} = 9.94 \times
                        10^{-11}T^{-0.6687}~\cm^3\sec^{-1},
\end{equation}
(e) HeII (free $\rightarrow 1^1S$)
\begin{equation}
    \alpha_{\HeII,1^1S} = 1.32\times 10^{-11} T^{-0.480}
                ~\cm^3\sec^{-1},
\end{equation}
(f) HeII (dielectronic recombination):
\begin{eqnarray}
    \xi_{\HeII}& = & 1.9\times 10^{-3} T^{-1.5}
                 \exp(-4.7\times 10^5/T) \nonumber\\
               & & \times [1+ 0.3 \exp(-9.4\times 10^4/T)]
                 ~\cm^3\sec^{-1},
\end{eqnarray}
(g) HeIII (free $\rightarrow n\ge 1$):
\begin{equation}
    \alpha_{\HeIII} = 3.36\times 10^{-10}T^{-1/2}
               \left(\frac{T}{10^3}\right)^{-0.2}
                   \left[ 1 +
                   \left(\frac{T}{4\times 10^6}\right)^{0.7}\right]^{-1}
                   ~\cm^3\sec^{-1},
\end{equation}
(h) HeIII (free $\rightarrow m\ge 2$):
\begin{equation}
   \alpha_{\HeIII,2} = \left\{\begin{array}{ll}
        \alpha_{\HeIII} (1.11 - 0.044 \ln T) &
                  ~~~~T<2.2\times 10^4\\
        \alpha_{\HeIII} (1.43 - 0.076 \ln T) &
                  ~~~~T>2.2\times 10^4
        \end{array}\right. ,
\end{equation}
\subsection{Photoionization cross sections}
(a) HI ($n=1 \rightarrow $ free):
\begin{equation}
    \sigma_{\rm bf,\HI} = 1.18\times 10^{-11}\eph^{-4}
               \frac{e^{-4(\arctan z_1)/z_1}}{
                 1-e^{-2\pi/z_1}} ~\cm^2
\end{equation}
where $z_1 = [\eph/\epsilon_{\HI} -1]^{1/2}$.

\noindent
(b) HI ($n=2 \rightarrow$ free):
\begin{equation}
   \sigma_{\rm bf,\HI,2} = 1.08\times 10^{-13}\eph^{-3}
           \frac{(3+4z^2)(5+4z^2)e^{-4(\arctan(2z))/z}}{
                    (1+4z^2)^3 (1-e^{-2\pi/ z})}~\cm^2,
\end{equation}
where $z = [\eph/\epsilon_{\HI} -1/4]^2$.

\noindent
(c) HeI:
\begin{equation}
   \sigma_{\rm bf,\HeI} = 1.13 \times 10^{-14}
           \left(\frac{1}{\eph^{2.05}}-
                 \frac{9.775}{\eph^{3.05}}\right) ~\cm^2,
\end{equation}
(d) HeII:
\begin{equation}
    \sigma_{bf,\HeII} = 7.55\times 10^{-10}\eph^{-4}
               \frac{e^{-4(\arctan z_2)/z_2}}{
                 1-e^{-2\pi/z_2}} ~\cm^2
\end{equation}
where $z_2 = [\eph/\epsilon_{\HeII} -1]^{1/2}$.

\subsection{Cooling Rates}

\subsubsection{Collisional ionization cooling}

\noindent
(a) HI:
\begin{equation}
    \zeta_{\HI} = 1.27\times 10^{-21}T^{1/2}
               \left[ 1+ \left(\frac{T}{10^5}\right)^{1/2}\right]^{-1}
                  \exp(-1.58\times 10^5/T)~\erg~ \cm^3\sec^{-1},
\end{equation}
(b) HeI:
\begin{equation}
    \zeta_{\HeI} = 9.38\times 10^{-22}T^{1/2}
                  \left[ 1+ \left(\frac{T}{10^5}\right)^{1/2}\right]^{-1}
                  \exp(-2.85\times 10^5/T)~ \erg~\cm^3\sec^{-1},
\end{equation}
(c) HeI($2^3S$):
\begin{eqnarray}
    \zeta_{\HeI(2^3S)}n(\HeI) & = & 5.01\times 10^{-27}T^{-0.1687}
                  \left[ 1+ \left(\frac{T}{10^5}\right)^{1/2}\right]^{-1}
                  \nonumber\\
                   &\times &\exp(-5.53\times 10^4/T)
                  n_en(\HeII)~\erg~\sec^{-1}.
\end{eqnarray}
(d) HeII:
\begin{equation}
    \zeta_{\HeII} = 4.95\times 10^{-22}T^{1/2}
                  \left[ 1+ \left(\frac{T}{10^5}\right)^{1/2}\right]^{-1}
                  \exp(-6.31\times 10^5/T)~ \erg~\cm^3\sec^{-1},
\end{equation}
\subsubsection{Recombination cooling}
\noindent
(a) HII:
\begin{equation}
   \eta_{\HII} = 6.50\times
           10^{-27}T^{1/2}\left(\frac{T}{10^3}\right)^{-0.2}
           \left[1+\left(\frac{T}{10^6}\right)^{0.7}\right]^{-1}
           ~\erg~\cm^3\sec^{-1},
\end{equation}
(b) HeII:
\begin{equation}
    \eta_{\HeII} = 1.55\times 10^{-26} T^{0.3647}~\erg~\cm^{-3}\sec^{-1},
\end{equation}
(c) HeII (dielectronic recombination)
\begin{eqnarray}
    \omega_{\HeII} &= & 1.24\times 10^{-13}T^{-1.5}
                \exp(-4.7\times 10^5/T)\nonumber\\
           & & \times
                [1+0.3\exp(-9.4\times 10^4/T)]
                ~\erg~\cm^3\sec^{-1},
\end{eqnarray}
(d) HeIII:
\begin{equation}
   \eta_{\HeIII} = 3.48\times
           10^{-26}T^{1/2}\left(\frac{T}{10^3}\right)^{-0.2}
           \left[1+\left(\frac{T}{4\times 10^6}\right)^{0.7}\right]^{-1}
           ~\erg~\cm^3\sec^{-1}.
\end{equation}

\subsubsection{Collisional excitation cooling}
\noindent
(a) HI:
\begin{equation}
     \psi_{\HI} = 7.5\times 10^{-19}
                  \left[ 1+ \left(\frac{T}{10^5}\right)^{1/2}\right]^{-1}
                  \exp(-1.18\times 10^5/T)~ \erg~\cm^3\sec^{-1},
\end{equation}
(b) HeI
\begin{eqnarray}
    \psi_{\HeI}n(\HeI) & = & 9.10\times 10^{-27}T^{-0.1687}
                  \left[ 1+
                  \left(\frac{T}{10^5}\right)^{1/2}\right]^{-1}
                  \nonumber\\
               & \times &
                  \exp(-1.31\times 10^4/T)n_en(\HeII)~ \erg~\sec^{-1},
\end{eqnarray}
(c) HeII:
\begin{equation}
     \psi_{\HeII} = 5.54\times 10^{-17}T^{-0.397}
                  \left[ 1+ \left(\frac{T}{10^5}\right)^{1/2}\right]^{-1}
                  \exp(-4.73\times 10^5/T)~ \erg~\cm^3\sec^{-1}.
\end{equation}

\subsubsection{Free-free cooling}
\begin{equation}
    \theta_{\rm ff} = 1.42\times^{-27}g_{\rm ff}T^{1/2}.
\end{equation}

\subsubsection{Compton cooling}
\begin{equation}
   \lambda_{\rm c} = 4k(T_e-T_{\gamma})\frac{\pi^2}{15}
           \left(\frac{kT}{\hbar c}\right)^3
           \left(\frac{kT}{m_e c^2}\right)n_e\sigma_{\rm T} c
\end{equation}

\newpage

\noindent
{\large\bf References}

\begin{description}
\item[] Adams, F., Bond, J.R, Freeses, K. Freeman, J., \& Olinto, A. 1992,
Phys. Rev. D47, 426.
\item[] Bajtlik, S., Duncan, R.C. \& Ostriker, J.P., 1988, ApJ
327, 570.
\item[] Barcons, X., Fabian, A.C. \& Rees, M.J. 1991, Nat 350,
685.
\item[] Bardeen, J.M., Bond, J.R., Kaiser, N. \& Szalay, A.S.  1986,
ApJ 304, 15.
\item[] Bartlett, J.G. \& Stebbins, A.  1991, ApJ 371, 8.
\item[] Binggeli, B., Sandage, A. \& Tammann, G. A. 1985, AJ  90, 1681.
\item[] Black, J.H. 1981, MNRAS 197, 553.
\item[] Blumenthal, G.R., Faber, S.M., Primack, J.R. \& Rees, M.J. 1984, Nat
311, 517.
\item[] Bond, J.R., Carr, B.J. \& Hogan, C.J. 1986, ApJ  306, 428.
\item[] Bond, J.R. \& Szalay, A.S. 1983, ApJ  274, 443.
\item[] Carr, B.J., Bond, J.R. \& Arnett, W.D. 1984, ApJ
277, 445.
\item[] Cen, R. 1992, ApJS  78, 341.
\item[] Cen, R., Jameson, A., Liu, A. \& Ostriker, J.P. 1990, ApJ 362,
L41.
\item[] Cen, R. \& Ostriker, J.P. 1992, ApJ 399, 331.
\item[] Couchman, H.M.P. 1985, MNRAS  214,
137.
\item[] Couchman, H.M.P. \& Rees, M.J. 1986, MNRAS
214, 137.
\item[] Crittenden, R., Bond, J.R., Davis, R.L., Efstatiou, G. \&
Steinhardt, P.J. 1993, Phys. Rev. Lett. 71, 324.
\item[] Davis, M, Summers, F. \& Schlegel, D. 1992, Nat 359,
393.
\item[] Davis, R.L., Hodges, H.M., Smoot, G.F., Steinhardt \& Turner,
M.S. 1992, Phys. Rev. Lett. 69, 1856.
\item[] Doglov, A.D. \& Silk, J. 1992, Berkeley preprint.
\item[] Doroshkevich, Zeldovich, Ya. B. \& Novikov, I. D. 1967,
Astr. Zh. 44, 295 [Soviet Astron. -- AJ 11, 233].
\item[] Efstathiou, G., Bond, J. R. \& White, S. D. M. 1992,
MNRAS 258, 1p.
\item[] Efstathiou, G. \& Rees, M.J. 1988, MNRAS 230, 5p.
\item[] Fukugita, M. \& Kawasaki, M. 1990, ApJ 353, 384.
\item[] Fukugita, M. \& Kawasaki, M. 1993, ApJ 402, 58.
\item[] Fukugita, M., Okamura, S. \& Yasuda, N. 1993, ApJ  412, L13
\item[] Frenk, C.S., White, S.D.M., Davis, M. \& Efstathiou, G.
1988, ApJ  327, 507.
\item[] Gaier, T., et al. 1992, ApJ 398, L1.
\item[] Gnedin, N.Y. \& Ostriker, J.P. 1992, ApJ 400, 1.
\item[] G\'orski, K., Stompor, R. \& Juszkiewicz, R. 1992, YITP
preprint.
\item[] Gunn, J.E. \& Peterson, B.A. 1965, ApJ  142, 1633.
\item[] Hogan, C.J. 1980, MNRAS 192, 891.
\item[] Jenkins, E.B. \& Ostriker, J.P. 1991, ApJ  376, 33.
\item[] Jones, B.J.T. \& Wyse, R.F.G. 1985, A\&A  149, 144.
\item[] Klypin, A., Holtzman, J., Primack, J. \& Rogos, E. 1992, ApJ
in press.
\item[] Madau, P. 1992, ApJ 389, L1.
\item[] Mather, J.C., et al. 1993, COBE preprint.
\item[] Matsuda, T., Sato, H., and Takeda, H. 1971, Prog. Theor.
Phys. 46, 416.
\item[] Menzel, D.H. \& Pekeris, C.L. 1935, MNRAS 96, 77.
\item[] Miralda-Escud\'e, J. \& Ostriker, J.P. 1990, ApJ 350,
1.
\item[] Miralda-Escud\'e, J. \& Ostriker, J.P. 1992, ApJ 392,
15.
\item[] Mo, H. J., Miralda-escud\'e, J. \& Rees, M. 1993, Cambridge IoA
preprint.
\item[] Ostriker, J.P. \& Ikeuchi, S. 1983, ApJ  263, L63.
\item[] Peebles, P.J.E. 1968, ApJ 153, 1.
\item[] Peebles, P.J.E. 1982, ApJ 263, L1.
\item[] Peebles, P.J.E. 1993, Principles of Physical Cosmology
(Princeton University Press).
\item[] Press, W.H. \& Schechter, P.L. 1974, ApJ 181,425.
\item[] Rees, M.J. 1986, MNRAS 218, 25p.
\item[] Sandage, A., Binggeli, B. \& Tammann, G. A. 1985, ApJ
90, 1759
\item[] Sargent, W.L.W., Young, P.J., Boksenberg \& Tytler, D. 1980,
ApJS  69, 703.
\item[] Sasaki, S., Takahara, F. \& Suto, Y. 1993, preprint
YITP/U-93-02.
\item[] Sugiyama, N., Silk, J. \& Vittorio, N. 1993, ApJ in press.
\item[] Shapiro, S.L. \& Teukolsky, S.A. 1983, `Black Holes, White
Dwarfs, and Neutron Stars', (John Wiley and Sons) p9.
\item[] Smoot, G.F., et al. 1992, ApJ {  396}, L1.
\item[] Spitzer, L., Jr. 1978, Physical Processes in the Interstellar
Medium (New York: Wiley).
\item[] Stebbins, A. \& Silk, J. 1986, ApJ  300, 1.
\item[] Steidel, C.C. \& Sargent, W.L. 1989, ApJ  343, L33.
\item[] Yoshioka, S. \& Ikeuchi, S. 1987, ApJ 323, L7.
\item[] Walker, T.P., Steigman, G., Schramm, D.N., Olive, K.A. and
Kang, H.-S. 1991, ApJ  376, 51.
\item[] Webb, J.K., Barcons, X., Carswell, R.F. \& Parnell, H.C.
1992, MNRAS 255, 319.
\item[] White, S.D. \& Frenk, C.S. 1991, ApJ  379, 52.
\end{description}
\newpage

{\large\bf Figure Captions}

\begin{description}

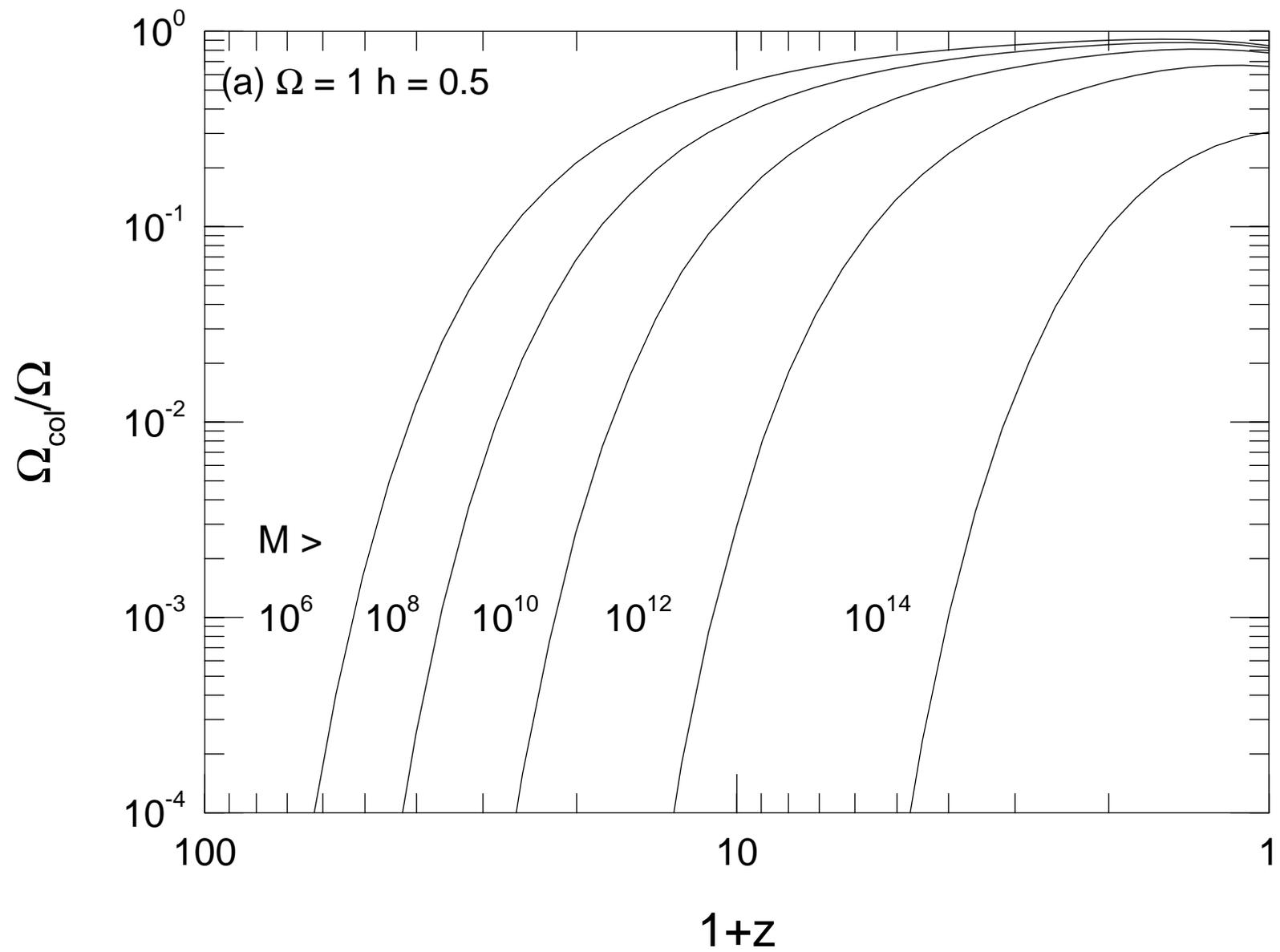
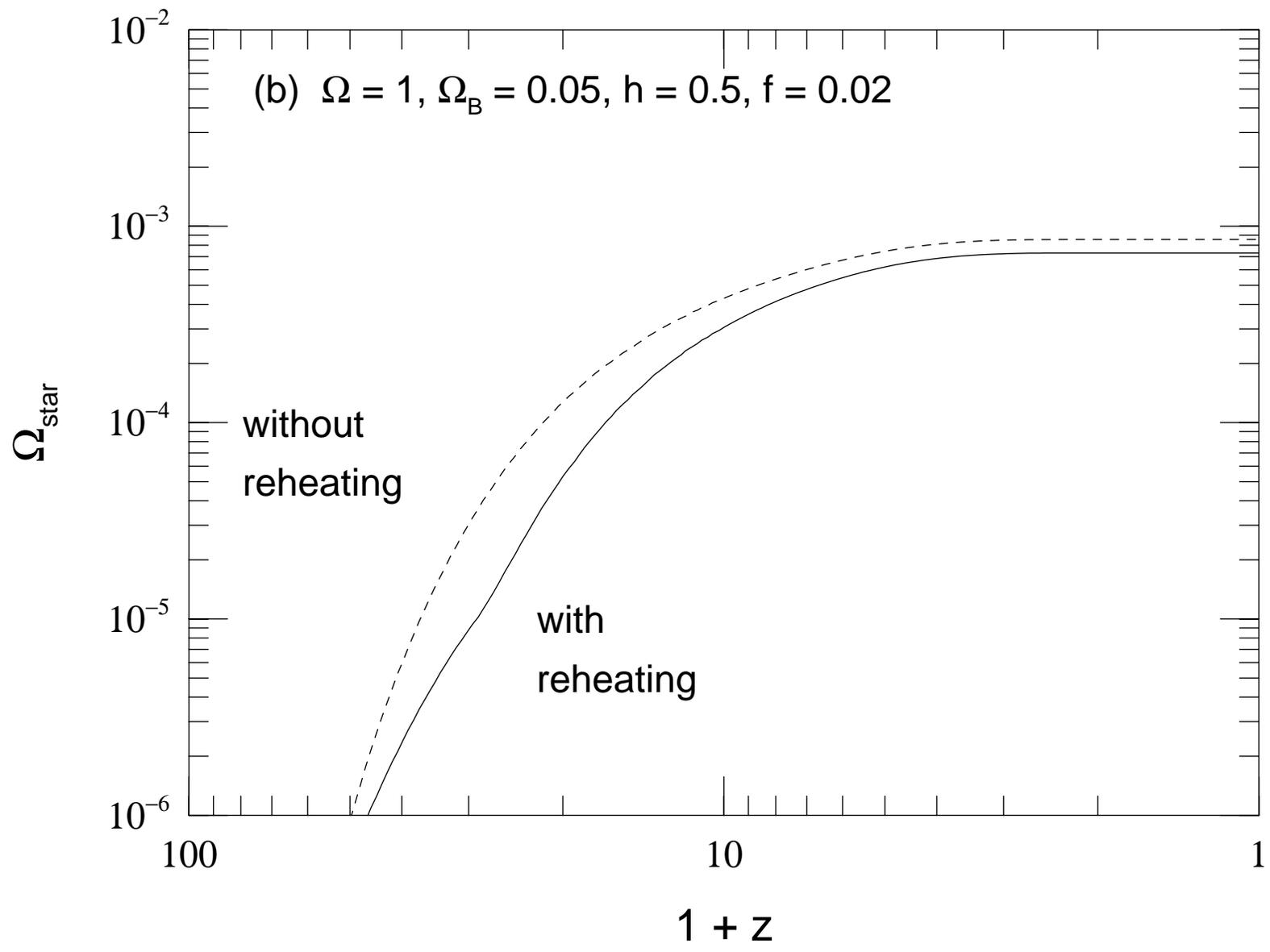
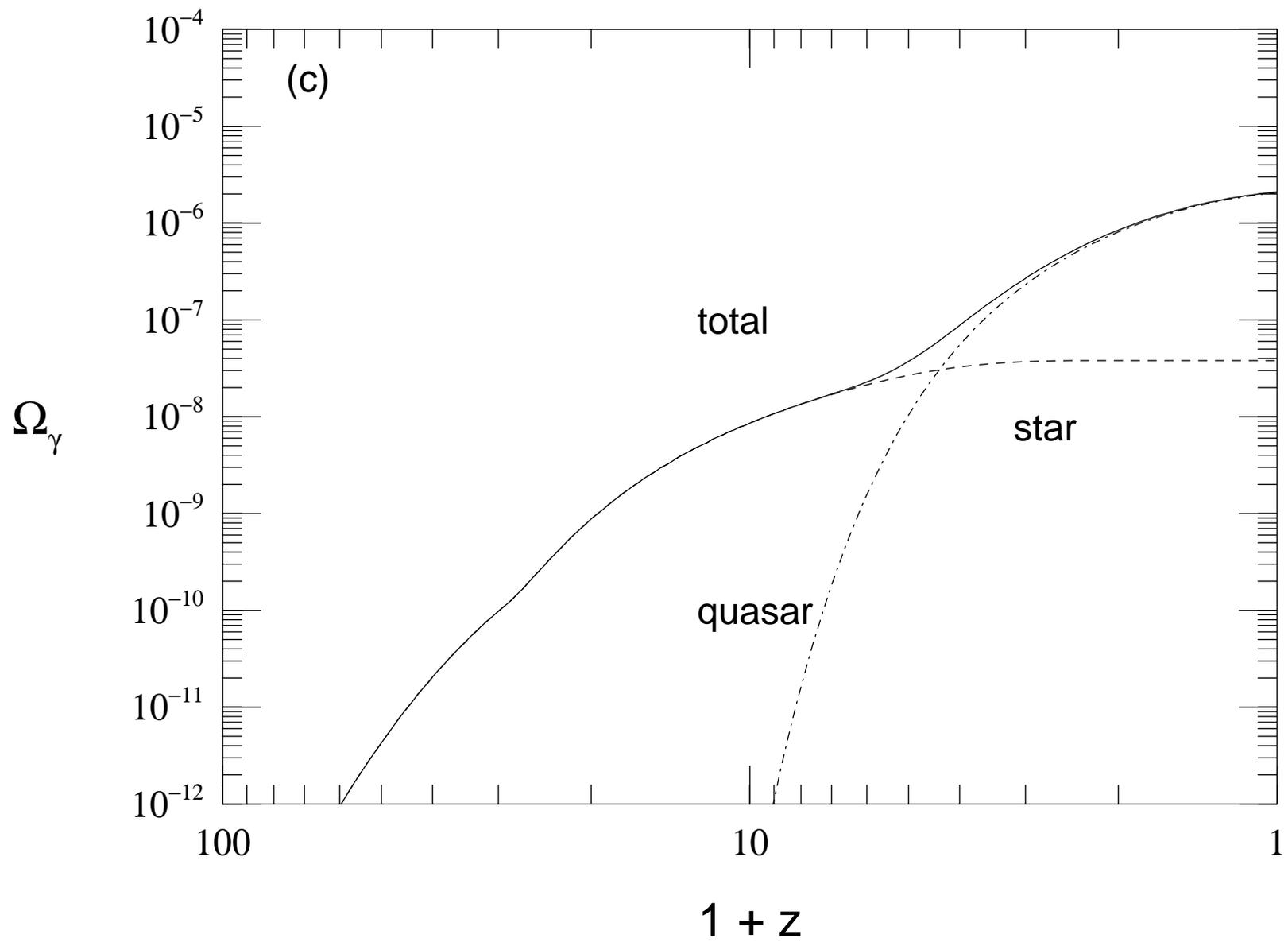
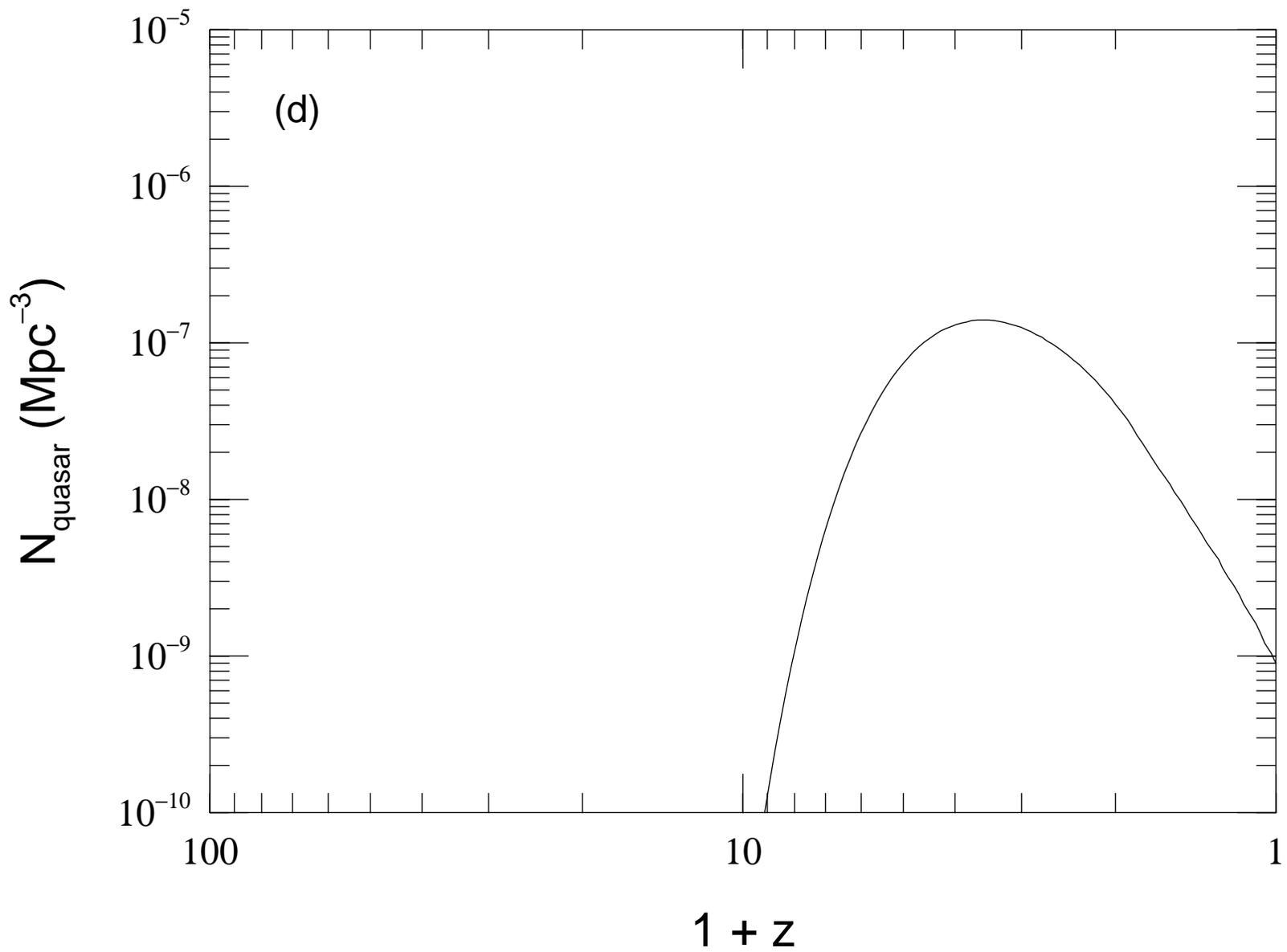
\item[Fig.1] (a) Evolution of the density of collapsed objects with
mass $ M$ greater than prescribed values (shown in units of
$M_{\odot}$)
for $\Omega =1$, $h=0.5$.\\
(b) Evolution of the mass fraction of stars for
$\Omega_B = 0.05$ and $f = 0.02$ ($\Omega$ and $h$ are the same as in
(a)) with the effect of
reionization taken into account (solid curve). The dashed curve represents
the mass fraction of stars when the effect of reionization is not
taken into account.\\
(c) Evolution of the radiation energy emitted from stars (dashed
curve) and from quasars (dashed-dotted curve). The solid
curve represents the sum. \\
(d) Comoving number density of quasars plotted against redshift for
$t_{\rm Q} = 10^8$year and $\epsilon_{\rm Q} F_{\rm Q} = 10^{-5}$.

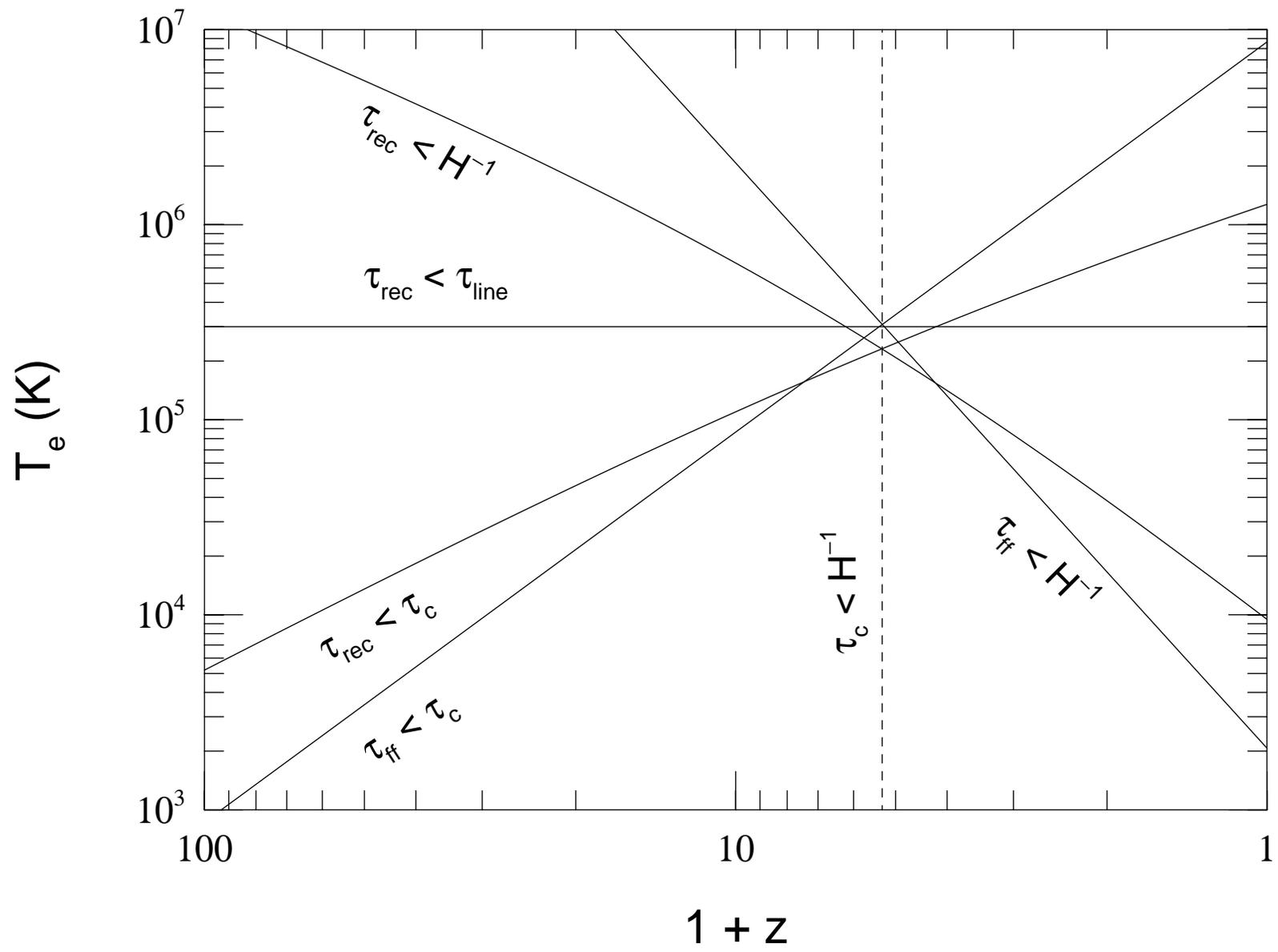
\item[Fig.2] Critical temperatures for $\tau_{\rm rec}
= \tau_{\rm C}$, $\tau_{\rm ff} = \tau_{\rm C}$, $\tau_{\rm rec}= H^{-1}$,
$\tau_{\rm ff}= H^{-1}$, $\tau_{\rm C} = H^{-1}$ plotted as a function of
redshift. Region for $\tau_{\rm rec}\le \tau_{\rm C}, H^{-1}$ or
$\tau_{\rm ff}\le \tau_{\rm C}, H^{-1}$ is needed for high energy
photon emission from the collapsing objects. Another requirement that
$\tau_{\rm rec}\le \tau_{\rm line}$, which is estimated from Fig.6
below, is also shown. There are no regions that satisfy all the
requirement.

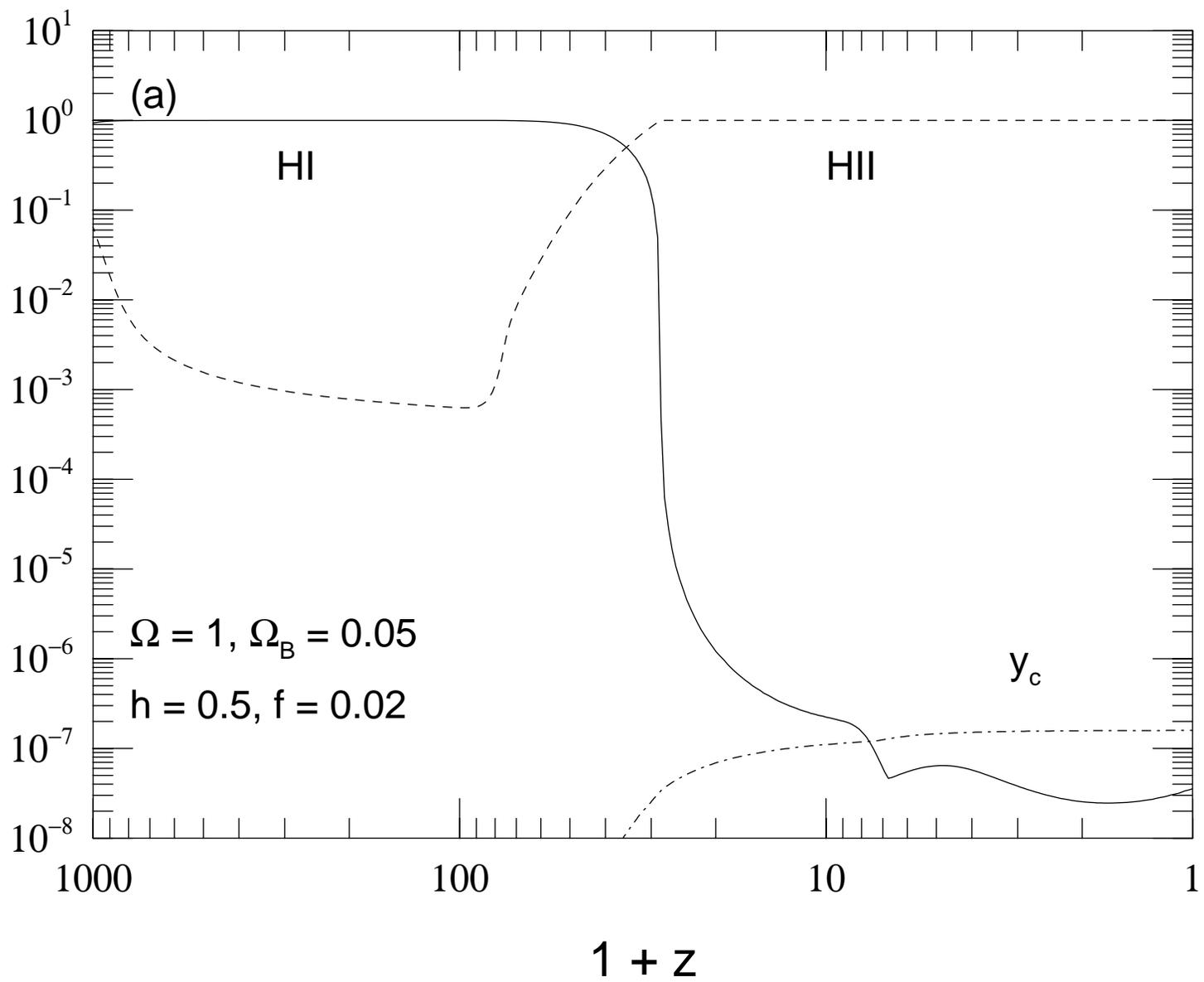
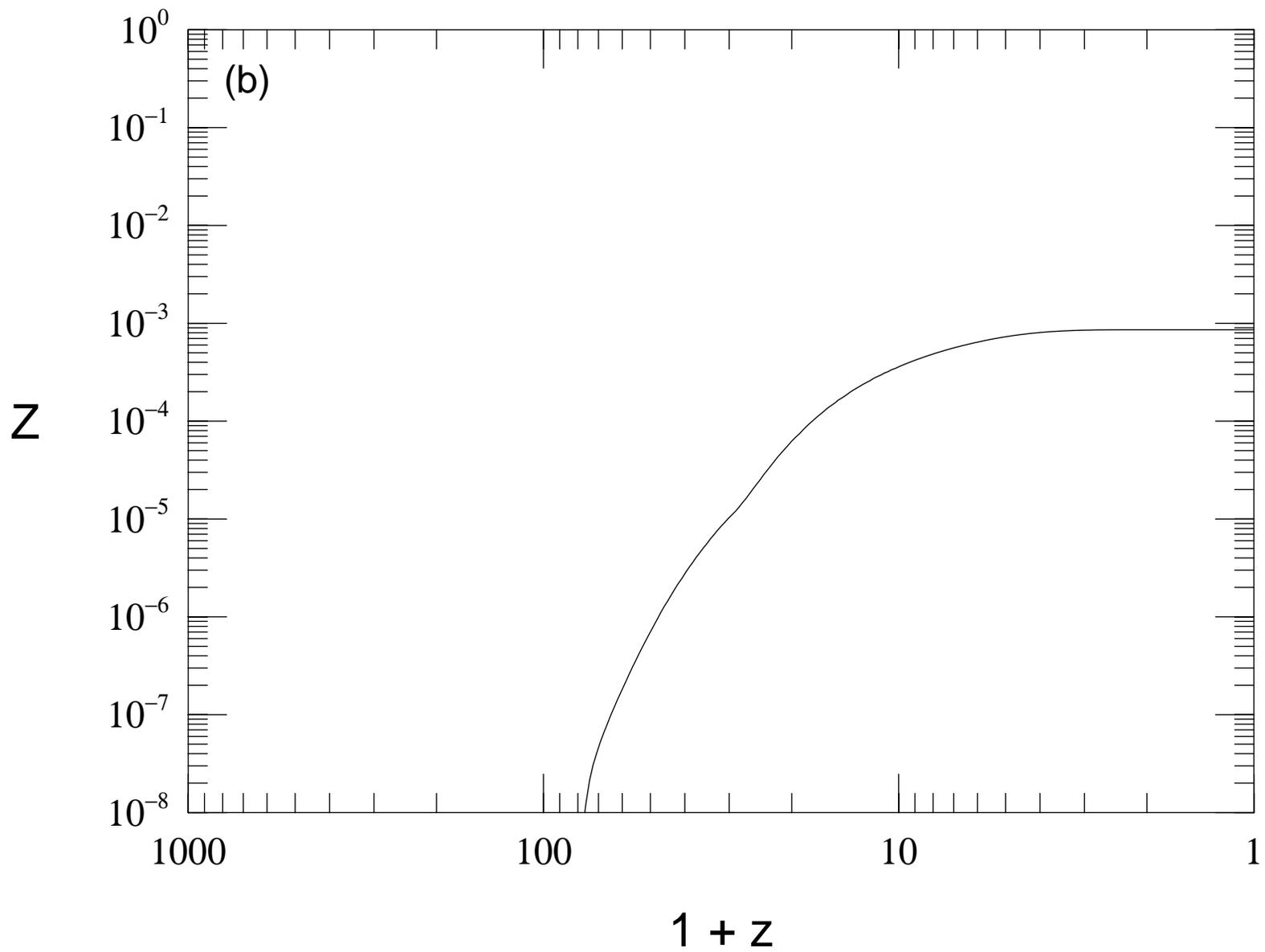
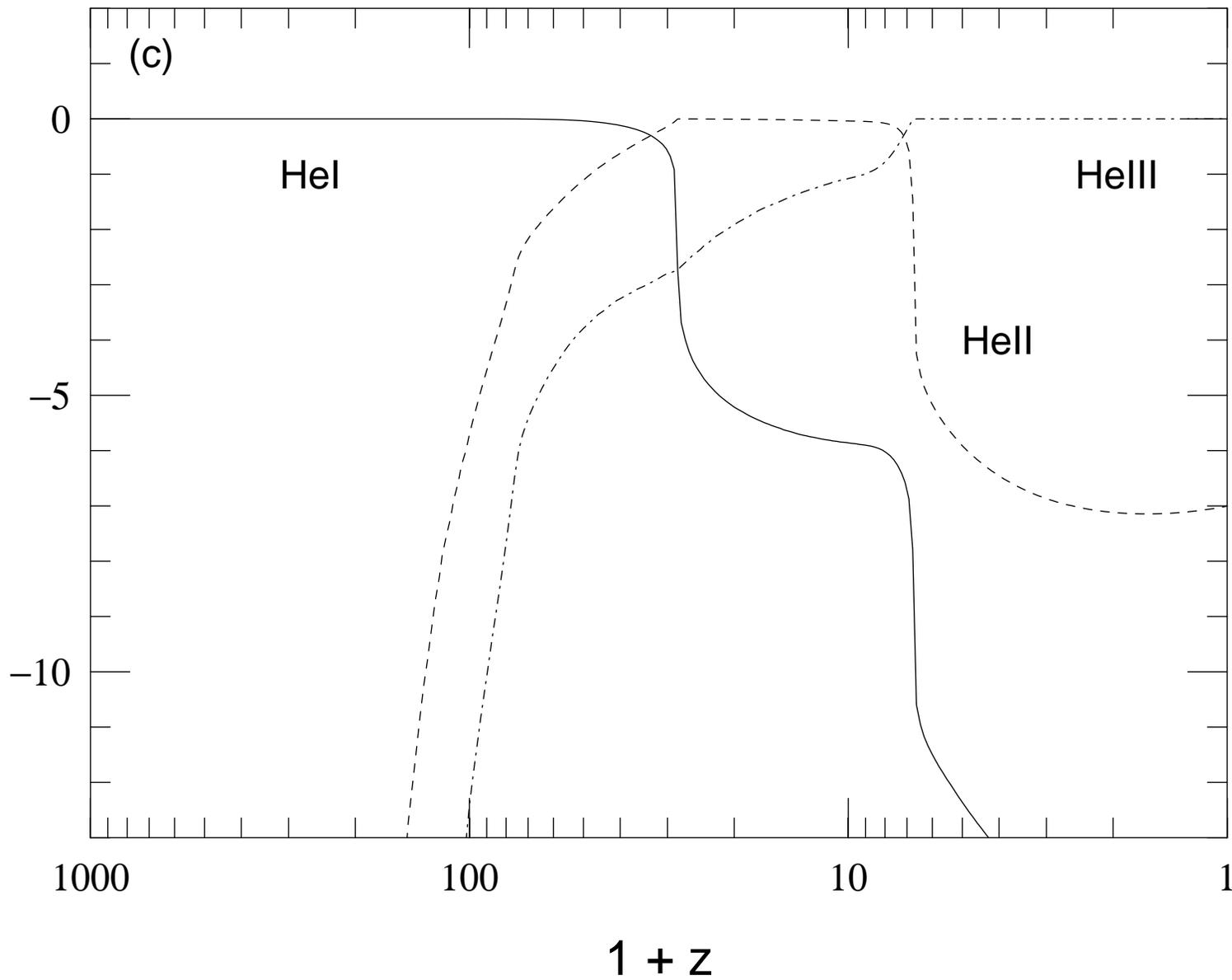
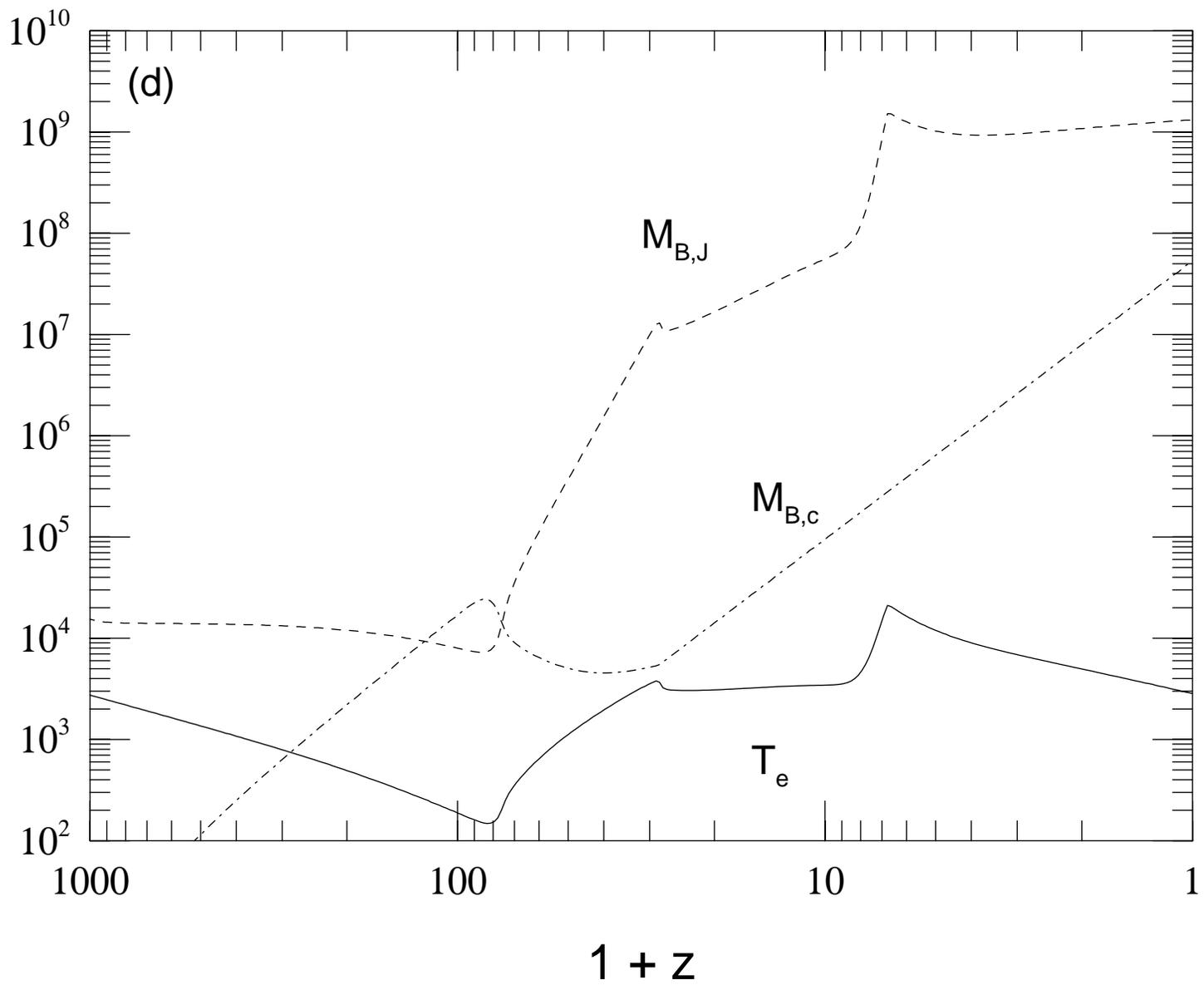
\item[Fig.3] (a) Evolution of the fraction of  HI (solid curve),
HII (dashed curve) and $y_{\rm c}$-parameter (dashed-dotted curve)
shown for $\Omega = 1$,
$\Omega_B = 0.05$, $h=0.5$ and $f=0.02$.\\
(b) Evolution of metalicity $Z$ as a function of redshift. \\
(c) Evolution of the fraction of HeI (solid curve), HeII(dashed
curve) and HeIII (dashed-dotted curve).\\
(d) Evolution of electron temperature $T_e$(K), baryonic Jeans mass
$M_{B,J}$, and the critical mass $M_{B,c}$ above which the molecular
cooling time is shorter than the dyanmical time. $M_{B,J}$ and $M_{B,c}$
are in units of $M_{\odot}$.

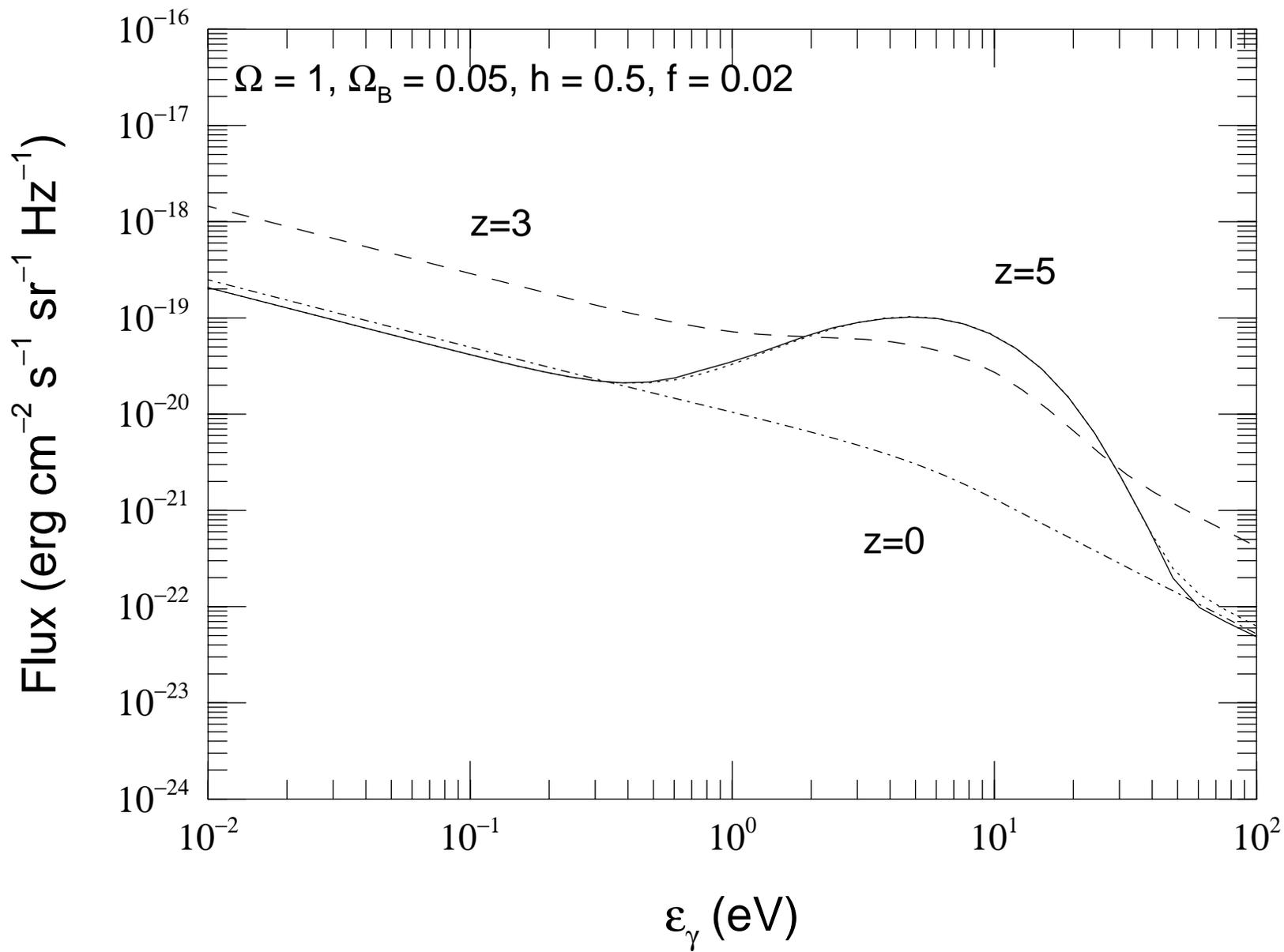
\item[Fig.4] Residual photon spectra at $z=5$ (solid curve), $z=3$
(dashed curve) and $z=0$ (dashed-dotted curve)  for $\Omega = 1$,
$\Omega_B = 0.05$, $h=0.5$ and $f=0.02$.
The dotted curve that closely follows the curve for $z=5$ stands for
the spectrum ($z=5$) when the effect of absorption due to
intergalactic  H and He would be switched off.

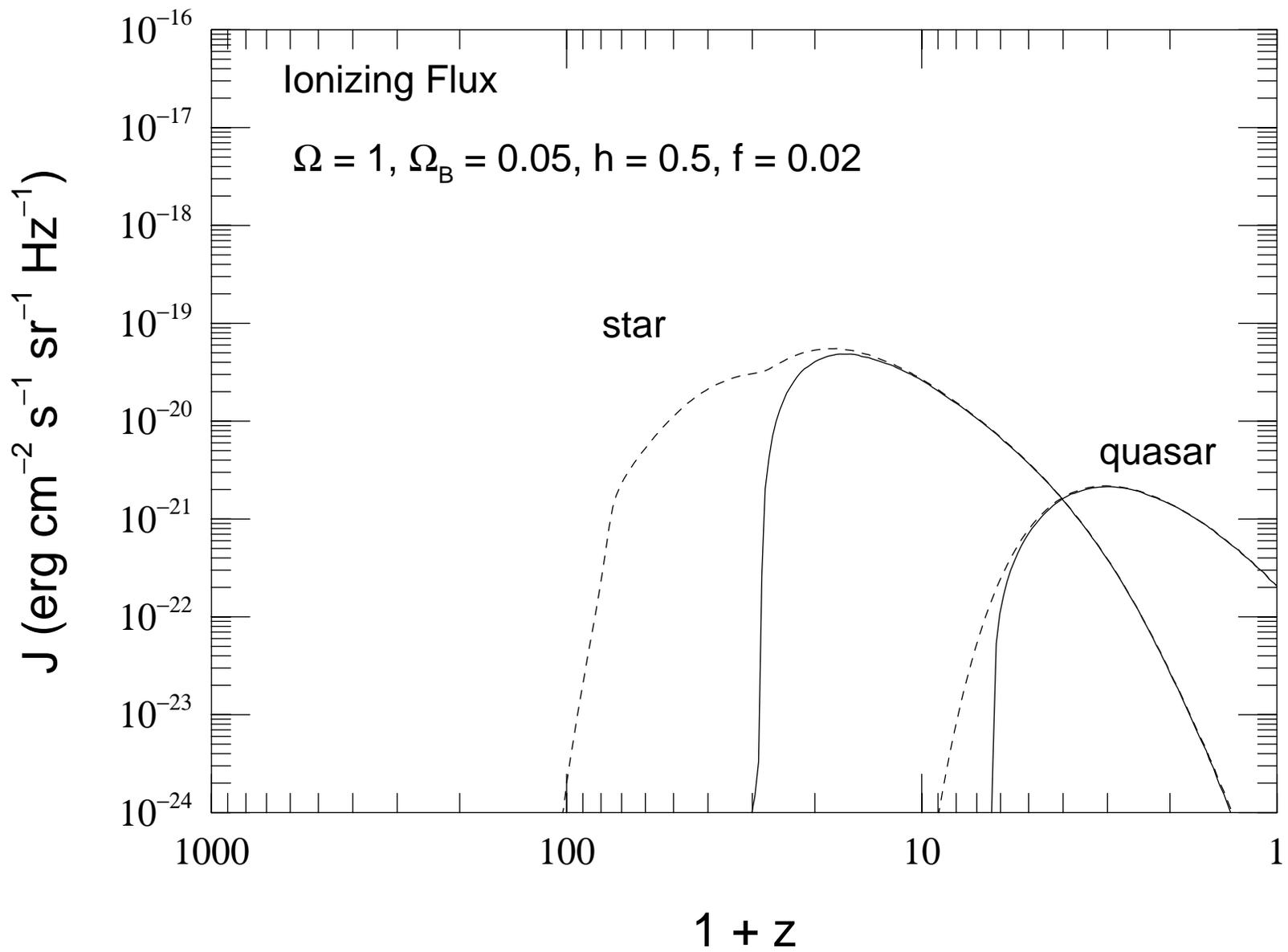
\item[Fig.5] Evolution of the ionizing flux from stars and quasars shown
for $\Omega = 1$, $\Omega_B = 0.05$, $h=0.5$ and $f=0.02$ (solid curves).
Dashed curves represent the ionizing fluxes which are obtained when
the effect of absorption due to H and He is switched off.

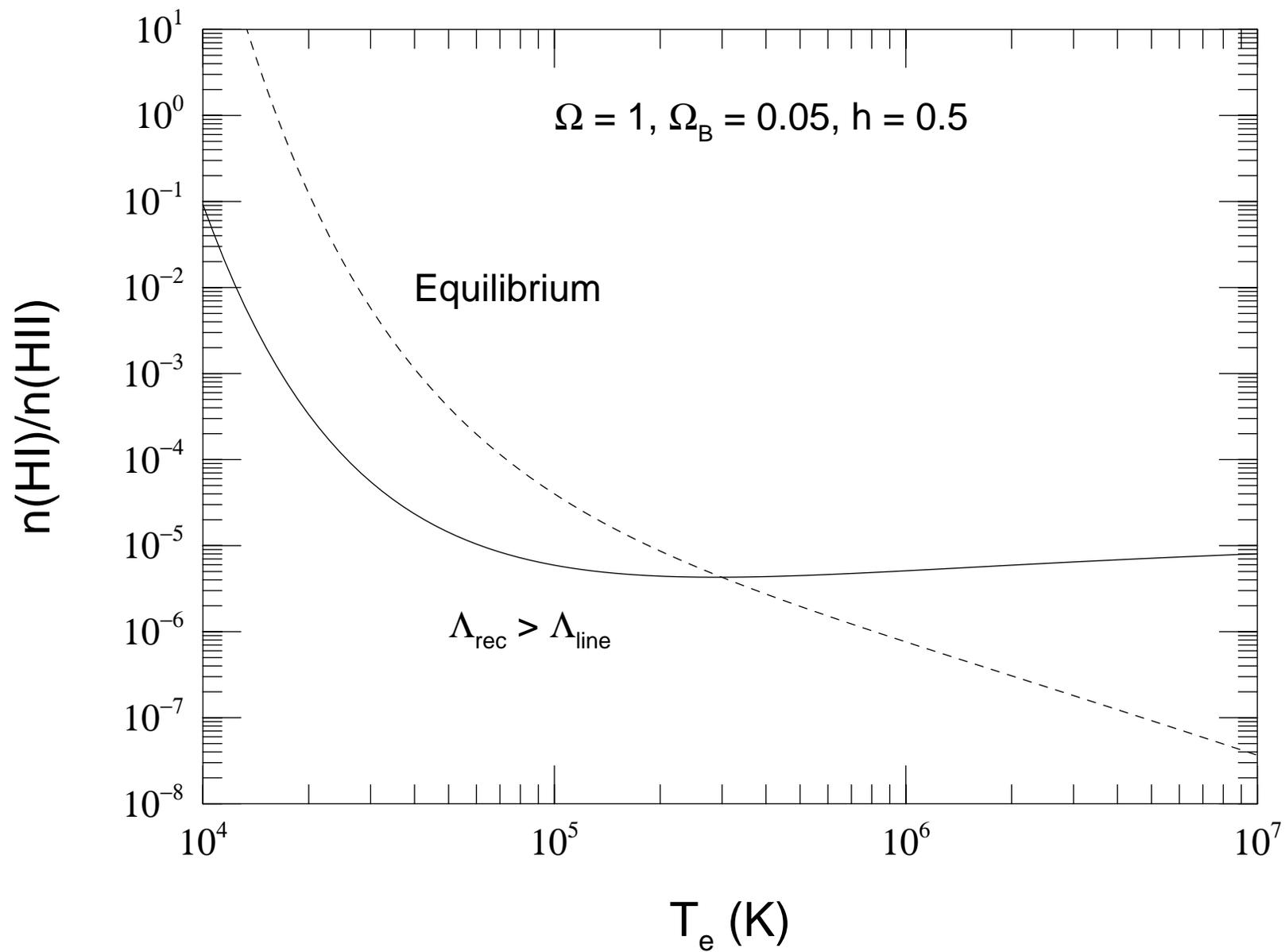
\item[Fig.6] Condition for recombination cooling to dominate over line
cooling for $\Omega = 1$,
$\Omega_B = 0.05$ and $h=0.5$. The region below the solid line shows
the parameters for which
recombination cooling dominates.
The dashed curve shows the ratio $n(\HI)/n(\HII)$ from
ionization equilibrium.

\end{description}

\end{document}